\documentclass[a4paper,fleqn]{cas-dc}

\usepackage[numbers]{natbib}
\usepackage{ulem}

\def\tsc#1{\csdef{#1}{\textsc{\lowercase{#1}}\xspace}}
\tsc{WGM}
\tsc{QE}
\tsc{EP}
\tsc{PMS}
\tsc{BEC}
\tsc{DE}
\usepackage{amsmath,amssymb,amsfonts}
\usepackage{textcomp}
\usepackage{xcolor}
\usepackage{hyperref}
\usepackage{graphicx}
\usepackage{geometry}
\usepackage{booktabs}
\usepackage{array}
\usepackage{multirow}
\usepackage{lscape}
\usepackage{tcolorbox}
\usepackage{rotating}
\usepackage{parskip}
\usepackage{stfloats}
\usepackage{subcaption}
\usepackage{enumitem}
\usepackage[utf8]{inputenc}
\usepackage{newunicodechar}
\newunicodechar{✱}{*}

\usepackage{awesomebox} 
\usepackage{fontawesome5}
\usepackage{hhline}
\definecolor{findingBarColour}{gray}{0.4}
\definecolor{findingIconColour}{gray}{0.3}

\usepackage{tcolorbox}

\newtcolorbox{rqbox}[1]
{
  before skip=1em, 
  after skip=1em, 
  colframe = black!60,
  colback  = black!10,
  coltitle = black!90,  
  title    = \textbf{#1},
  hbox boxed title,
  enhanced,
  attach boxed title to top center={yshift=-3mm,yshifttext=-1mm},
  boxed title style={size=small, colback=black!30}
}

\usepackage{siunitx}
\newcommand{\commentout}[1]{}

\usepackage{booktabs}

\usepackage{amssymb}
\usepackage{graphicx}
\usepackage{textcomp}
\usepackage{xcolor}
\usepackage{subcaption}
\usepackage{colortbl}
\usepackage{enumerate}
\usepackage{algpseudocode}

\usepackage{color}
\usepackage{hyperref}
\usepackage{xcolor}

\usepackage{booktabs}
\usepackage{array}

\usepackage[utf8]{inputenc}

\usepackage{adjustbox} 

\usepackage{booktabs}
\usepackage{amsmath}
\usepackage{array}

\usepackage{tcolorbox}

\begin{document}
\let\WriteBookmarks\relax
\def\floatpagepagefraction{1}
\def\textpagefraction{.001}

\shorttitle{Challenges in Test Mocking: Insights from StackOverflow}

\shortauthors{Ahmed et al.}

\title [mode = title]{Exploring Challenges in Test Mocking: Developer Questions and Insights from StackOverflow}                      




%

\author[1]{Mumtahina Ahmed}[]

\ead{mumtahina.ahmed@usask.ca}

\affiliation[1]{organization={SR Lab, Department of Computer Science, University of Saskatchewan},
    city={Saskatoon},
    country={Canada}}

\affiliation[2]{organization={SQM Research Lab, Department of Computer Science, University of Manitoba},
    city={Winnipeg},
    country={Canada}}

\author[2]{Md Nahidul Islam Opu}


\ead{opumni@myumanitoba.ca}

\author[1]{Chanchal Roy}[
   ]
\ead{chanchal.roy@usask.ca}
\author[2]{Sujana Islam Suhi}[
   ]
\ead{suhisi@myumanitoba.ca}

\author[2]{Shaiful Chowdhury}[
   ]
\ead{shaiful.chowdhury@umanitoba.ca}

\begin{abstract}
Mocking is a common unit testing technique that is used to simplify tests, reduce flakiness, and improve coverage by replacing real dependencies with simplified implementations. Despite its widespread use in Open Source Software (OSS) projects, there is limited understanding of how and why developers use mocks and the challenges they face. In this collaborative study, we have analyzed 25,302 questions related to \textit{Mocking} on \textsc{Stackoverflow} to identify the challenges faced by developers. We have used Latent Dirichlet Allocation (LDA) for topic modeling, identified 30 key topics, and grouped the topics into five key categories. Consequently, we analyzed the annual and relative probabilities of each category to understand the evolution of mocking-related discussions. Trend analysis reveals that category like \textit{Advanced Programming} peaked between 2009 and 2012 but have since declined, while categories such as \textit{Mocking Techniques} and \textit{External Services} have remained consistently dominant, highlighting evolving developer priorities and ongoing technical challenges.

Our findings also show an inverse relationship between a topic's popularity and its difficulty. Popular topics like \textit{Framework Selection} tend to have lower difficulty and faster resolution times, while complex topics like \textit{HTTP Requests and Responses} are more likely to remain unanswered and take longer to resolve. Additionally, we evaluated questions based on the answer status\textemdash{}successful, ordinary, or unsuccessful, and found that topics such as \textit{Framework Selection} have higher success rates, whereas tool setup and Android-related issues are more often unresolved. A classification of questions into \textit{How}, \textit{Why}, \textit{What}, and \textit{Other} revealed that over 70\% are \textit{How} questions, particularly in practical domains like file access and APIs, indicating a strong need for implementation guidance. \textit{Why} questions are more prevalent in error-handling contexts, reflecting conceptual challenges in debugging, while \textit{What} questions are rare and mostly tied to theoretical discussions. These insights offer valuable guidance for improving developer support, tooling, and educational content in the context of mocking and unit testing. 
\end{abstract}




\maketitle


\section{Introduction}
\label{sec:intro}
\normalem
In software testing, units often depend on other components~\cite{runeson2006survey}. Developers must decide whether to test a unit with its dependencies (similar to integration testing) or simulate them to test the unit in isolation. Testing with dependencies increases realism, as it better reflects production behavior~\cite{weyuker1998testing}. However, dependencies such as databases and web services can slow tests~\cite{samimi2013declarative}, be costly to set up~\cite{samimi2013declarative} and require complete control over external systems~\cite{freeman2009growing}. Simulating dependencies focuses the test on the unit itself and mitigates these inefficiencies. Mocking frameworks, such as Mockito~\cite{mockito}, EasyMock~\cite{easymock}, and JMock~\cite{jmock} for Java, and Mock~\cite{mock} and Mocker~\cite{mocker} for Python, support this by allowing developers to create simulated objects, set method return values, and verify interactions. Research shows the widespread use of mocking frameworks~\cite{henderson2017software} and suggests that they facilitate unit testing. To evaluate their impact and support practitioners, it is essential to understand and quantify developers' practices, which can guide future research, tool development, and testing processes.

While mocking is crucial for unit testing, selecting appropriate classes to mock remains challenging~\cite{mackinnon2000endo, spadini2019mock}. Differences in mocking frameworks, the debate over their benefits and drawbacks~\cite{freeman2009growing, spadini2019mock}, and the difficulties that automated tools face in making context-aware recommendations~\cite{arcuri2014automated} highlight the need for further investigation. Despite the extensive literature on mock implementation in various languages~\cite{freeman2009growing, hamill2004unit, meszaros2007xunit}, the practical use, challenges, and evolution of mocking remain unclear, making \textsc{Stackoverflow} (SO) a valuable resource for insight. Prior research has identified major developer concerns on \textsc{Stackoverflow} using topic modeling and quantitative analysis. Studies have applied NLP-based topic modeling to uncover key topics, trends, and challenges in developer Q\&A~\cite{haque2020challenges, bagherzadeh2019going, uddin2021empirical, abdellatif2020challenges}. In this context, examining discussions related to mocking can provide deeper insights into the specific obstacles and practices developers encounter in testing workflows.

In light of these limitations, we conduct the first in-depth study on the challenges developers face when using mocking frameworks, leveraging discussions from \textsc{Stackoverflow}. Mocking frameworks are widely used and play a crucial role in unit testing, yet developers often encounter difficulties in their implementation~\cite{mostafa2014empirical}. Given \textsc{Stackoverflow}'s prominence as a Q\&A platform for software developers, we analyze mocking-related posts to identify key discussion topics, categorize these discussions, and gain insights into the challenges developers face. Specifically, our study addresses the following research questions:
\begin{itemize}
    \item \textbf{RQ1: What are the most common topics discussed about mocking on \textsc{Stackoverflow}?}\\
    Using topic modeling, we identified 30 unique topics that were organized into five primary categories: \textit{Mocking Techniques}, \textit{Advanced Programming}, \textit{External Services}, \textit{Error Handling}, and \textit{Theoretical} aspects in mocking-related \textsc{Stackoverflow} questions. We presented a hierarchy consisting of mocking categories, subcategories, and topics. We also characterized each topic using top keywords with post-frequency. \textit{Mocking Python Components} was the most discussed topic, while \textit{Tool Setup Errors} was less frequent.
    \item \textbf{RQ2: How have mocking-related discussions on \textsc{Stackoverflow} evolved over the years?}\\
    Our investigation into the evolution of mocking-related discussions on SO revealed an early prevalence of \textit{Theoretical} and \textit{Advanced Programming} categories, peaking before 2016, followed by steady growth in the other three categories, with \textit{Mocking Techniques} leading by 2024. Topic-wise, early discussions focused on \textit{Framework Selection}, whereas more recent years, particularly post-2018, saw increased attention to \textit{JS Modules and Components}, and \textit{HTTP Requests and Responses}, reflecting a shift toward modern API-driven development. Although interest in \textit{Advanced Method Mocking} declined after 2015, topics like \textit{Exception Handling} and \textit{Database Queries} remained consistently relevant throughout the period.
    \item \textbf{RQ3: How do mocking-related topics vary in terms of popularity and difficulty?}\\
    We examined the popularity of mocking-related topics on \textsc{Stackoverflow} using metrics such as question score, view count, number of comments, and answer count. Difficulty was assessed based on the percentage of unanswered questions and the median time to receive an accepted answer. Overall, mocking-related questions demonstrated higher popularity compared to general \textsc{Stackoverflow} posts. Among these, topics like \textit{Framework Selection} were both popular and quickly resolved, indicating relative simplicity. In contrast, \textit{HTTP Requests and Responses} had more unanswered posts and slower resolutions, reflecting greater complexity or the need for deeper expertise.
    \item \textbf{RQ4: What types of questions Do developers ask about mocking?}\\
    We are trying to understand the types of problem developers face, so inspired by previous research, we categorized the posts into \textit{How}, \textit{Why}, \textit{What}, and \textit{Other} using automated labeling by GPT-4o-mini, followed by manual validation. The results show that most questions have a strong focus on implementation (\textit{How}) and troubleshooting (\textit{Why}), particularly in contexts involving external services and complex mocking scenarios. Compared to other software engineering fields, mocking-related inquiries have the highest proportion of practical (\textit{How}) questions, underscoring its implementation-driven nature. 
\end{itemize}
\subsection{Paper Organization}  
The remainder of this paper is organized as follows. Section~\ref{sec:related_work} reviews related work relevant to our research. Section~\ref{sec:datacollection} describes the data collection process of our study. Section~\ref{sec:results} presents our approach, analysis, and results. Section~\ref{sec:discussion} discusses the implications of our findings. Section~\ref{sec:threats} presents potential threats to validity. Finally, Section~\ref{sec:conclusion} concludes the paper.




\section{Related Work}
This section reviews the relevant literature in two key areas: the use and implications of mocking frameworks in software testing, and empirical research based on \textsc{StackOverflow} data. Together, these domains of research provide a foundation for investigating developer challenges and behaviors related to mocking practices.
\label{sec:related_work}
\subsection{Studies on Mocking Frameworks:} With the growing reliance on automated testing, mocking frameworks have become essential to isolate software components by simulating dependencies. Several empirical studies have examined their adoption, usage patterns, and impact on testing practices.

Mostafa and Wang~\cite{mostafa2014empirical} et al. analyzed 5,000 open-source GitHub projects, finding that 23\% of the test code used mocking frameworks such as Mockito, EasyMock, JMock, and JMockit. Developers preferred to mock source code classes over library classes, indicating the selective use of mocks to manage dependencies. Spadini et al.~\cite{spadini2017mock} examined more than 2,000 test dependencies in open-source and industrial projects, revealing that developers frequently mock external resources, such as databases and web services, to mitigate complexity and performance overhead. However, they also noted that excessive mocking can obscure design flaws by creating inconsistencies with real implementations.

Xiao et al.~\cite{xiao2024empirical} studied Apache Software Foundation projects, reporting that 66\% of the Java projects relied on Mockito, EasyMock, or PowerMock. Mocking was more prevalent in large and newer projects, with fluctuating usages as testing strategies evolved. Expanding beyond Java, Almeida et al.~\cite{de2023mock} investigated mocking in multiple programming languages. They found that while Mockito, PHPUnit, Jest, and Mock were widely used in Java, PHP, JavaScript, and Python, respectively, the number of mocks was often correlated with the number of test files. However, extensive use of mocks did not necessarily improve test coverage. 

In a broader study, Spadini et al. ~\cite{spadini2019mock} inspected over 2,000 mock usages. Their survey of 100+ developers revealed that mocks are primarily used for infrastructure dependencies, while domain-specific classes are rarely mocked. Maintaining consistency between mock behavior and real implementations remains a key challenge, as mocks can introduce unintended coupling and require ongoing maintenance as production code evolves. These studies collectively highlight the benefits and challenges of mocking frameworks. While mocks aid in dependency management and testability, they also introduce maintenance complexities that developers must carefully navigate.

While prior studies have provided valuable insights into how developers use mocking frameworks\textemdash{}highlighting their prevalence, usage patterns, and impacts on testing practices, they have not thoroughly explored the specific challenges developers encounter when using these frameworks. This gap motivates our work, which aims to investigate the practical difficulties developers face, as reflected in their discussions on \textsc{StackOverflow}.

\subsection{Studies on StackOverflow:} 
\textsc{StackOverflow} serves as the primary platform within the Stack Exchange Network, underscoring its key role for software professionals. It hosts approximately $24$ million questions, $36$ million answers, and about $29$ million registered users\footnote{\url{https://stackexchange.com/sites?view=list\#users} last accessed: 12-May-2025}. This extensive knowledge base has a notable impact in multiple areas of software engineering research, including software architecture and design~\cite{bi2021mining}, social aspects of software engineering~\cite{gantayat2015synergy}, API documentation~\cite{treude2016augmenting}, machine learning-based classifiers~\cite{chowdhury2015mining}, and the recommendation system~\cite{diamantopoulos2015employing}, among others. 

Similar to our study, several works that use \textsc{StackOverflow} data have focused on analyzing questions and answers to extract insights into practitioner discussions on specific topics. Researchers have applied topic modeling and qualitative methods to analyze SO data~\cite{bajaj2014mining, barua2014developers, yang2016security, rosen2016mobile}. Barua et al.~\cite{barua2014developers} used Latent Dirichlet Allocation (LDA) to analyze \textsc{StackOverflow} discussions, uncovering key topics and evolving trends in developer interests like web/mobile development, Git, and MySQL. Similarly, Bajaj et al.~\cite{bajaj2014mining} analyzed \textsc{StackOverflow} questions using unsupervised learning and a custom ranking algorithm to identify common web development challenges, highlighting persistent DOM issues, declining browser concerns, and increased interest in mobile and HTML5 features.

Topic modeling, an NLP technique, identifies related words within textual datasets, helping researchers uncover hidden structures without prior knowledge of the content~\cite{chen2016survey, blei2003latent}. It has proven particularly effective in organizing unstructured data, constituting 80–85\% of the data found in software repositories~\cite{liu2016overview}. Building on this potential, several studies have applied topic modeling to analyze software engineering discussions on platforms like \textsc{StackOverflow} (SO) and GitHub. For instance, Bagherzadeh et al.~\cite{bagherzadeh2019going} explored the challenges of Big data development by analyzing SO posts, offering insights into developer interests, topic popularity, and the difficulty levels of discussed topics.

Similarly, Uddin et al.~\cite{uddin2021empirical} examined IoT-related discussions, identifying key thematic areas such as Hardware, Software, Network, and Tutorials, with secure messaging and script scheduling emerging as major concerns. Abdellatif et al.~\cite{abdellatif2020challenges} conducted a topic modeling analysis of SO discussions on chatbot development, highlighting key themes such as implementation guidelines, system integration, and natural language understanding models. Similarly, Haque et al.~\cite{haque2020challenges} analyzed over 113,000 Docker-related SO posts, uncovering recurring challenges in application development, configuration, and networking. Zahedi et al.~\cite{zahedi2020mining} analyzed 12,989 SO posts, revealing a trend towards increasing the specificity of questions, with CI/CD errors and concepts being the dominant topics, while unanswered questions remained as a challenge. 

Peruma et al.~\cite{peruma2022refactor} examined refactoring-related discussions in SO, providing information on common refactoring intentions and real-world adoption challenges. Rosen et al.~\cite{rosen2016mobile} analyzed mobile software engineering discussions by examining 13 million \textsc{StackOverflow} posts, identifying key challenges related to app distribution, mobile APIs, data management, and UI development. Similarly, Han et al.~\cite{han2020programmers} conducted a comparative study on deep learning frameworks, uncovering common discussion patterns and workflow stages in TensorFlow, PyTorch, and Theano. Bangash et al. conducted a study on \textsc{StackOverflow}'s machine learning-related posts using the SOTorrent dataset, identifying popular and underrepresented topics~\cite{bangash2019developers}. Complementing these focused studies, Tanzil et al. conducted a systematic mapping study on \textsc{StackOverflow} based research in software engineering, identifying key trends, impacts, and future research opportunities across 18 SE domains~\cite{tanzil2025systematic}.

Together, these studies highlight the diverse applications of topic modeling in software engineering. Although numerous studies have utilized \textsc{StackOverflow} data across various fields, we found that none have specifically addressed mocking-related discussions. This paper seeks to address this gap by adopting methodologies from relevant prior research to analyze the challenges encountered by developers in the software testing community.

\section{Data Collection}
\label{sec:datacollection}
\begin{figure*}[t]
    \centering
    \includegraphics[width=\textwidth,keepaspectratio]{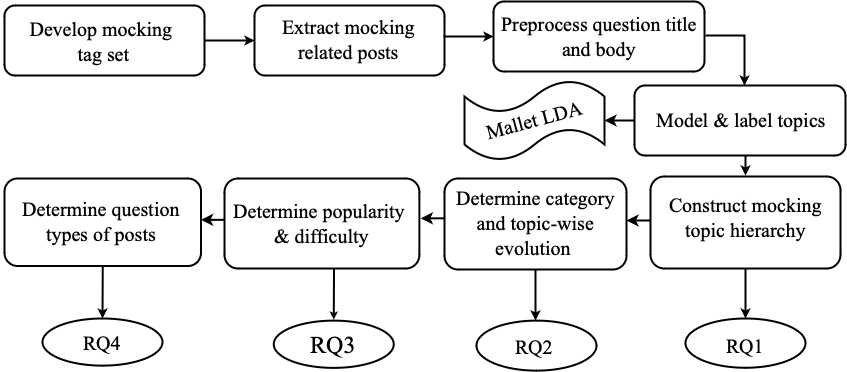} 
    \caption{Methodology of topic modeling on mocking-related posts}
    \label{fig:method}
\end{figure*}
\begin{figure*}[t]
    \centering

    \begin{subfigure}[t]{0.80\textwidth}
        \centering
        \includegraphics[width=\textwidth,keepaspectratio]{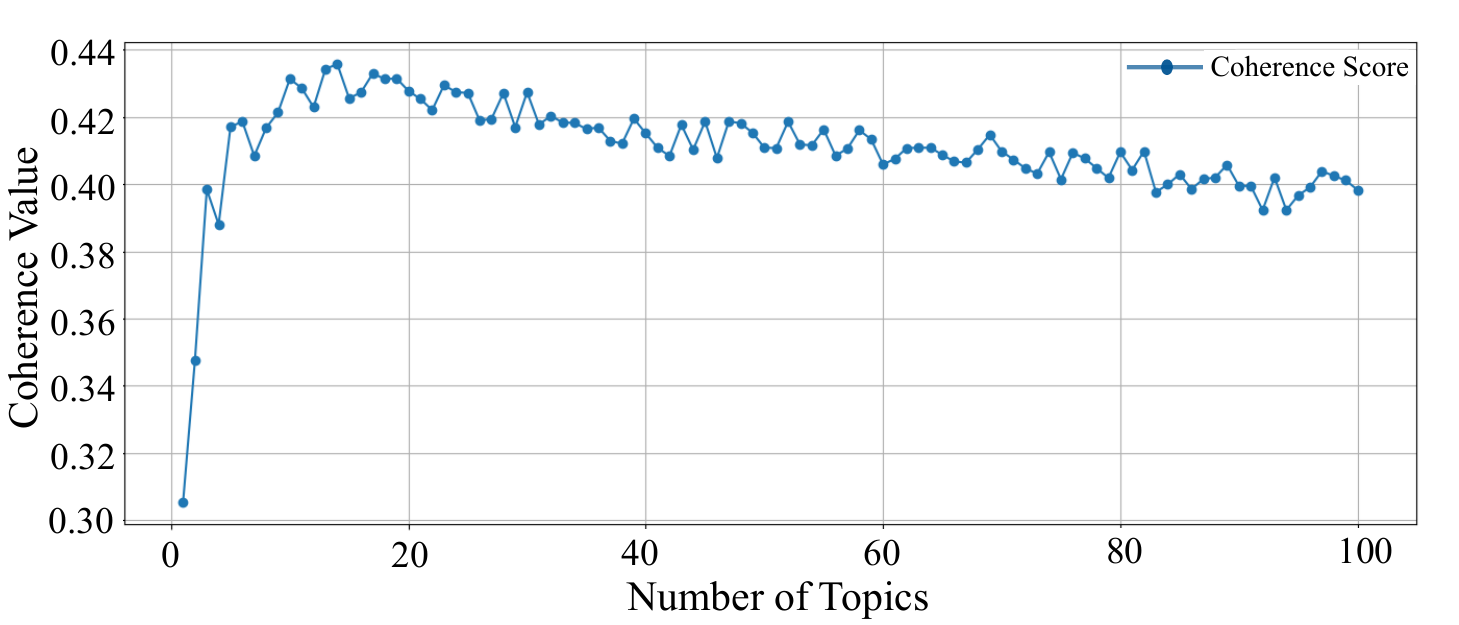}
        \caption{Tradeoff between Number of Topics and Coherence Value.}
        \label{fig:coherence-graph}
    \end{subfigure}

    \vspace{1em} 

    \begin{subfigure}[t]{0.95\textwidth}
        \centering
        \hspace*{1.5cm} \includegraphics[width=\textwidth,keepaspectratio]{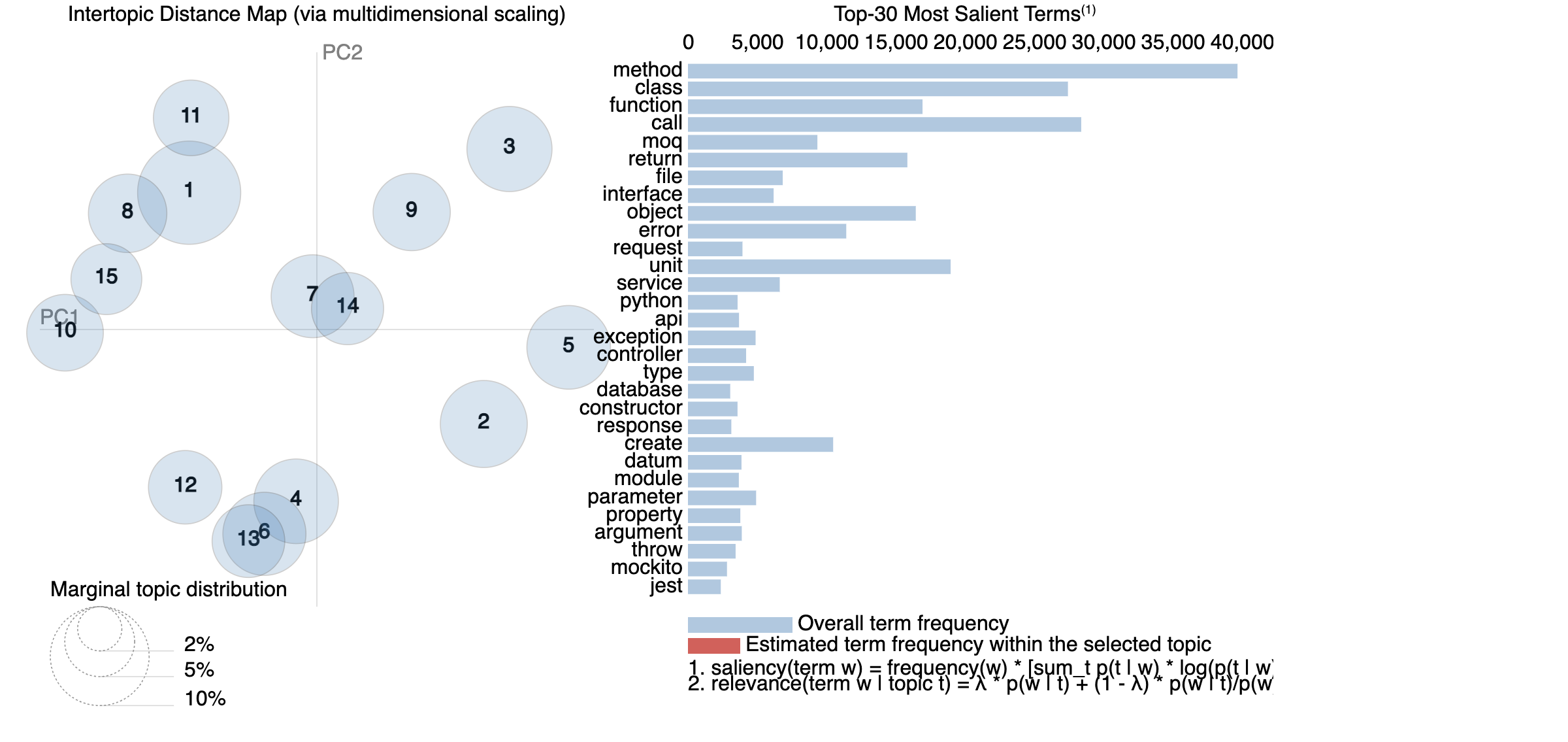}
        \caption{Topic visualization for 15-topic LDA model using pyLDAvis.}
        \label{fig:15Topics}
    \end{subfigure}

    \begin{subfigure}[t]{0.95\textwidth}
        \centering
        \hspace*{1.5cm}
         \includegraphics[width=\textwidth,keepaspectratio]{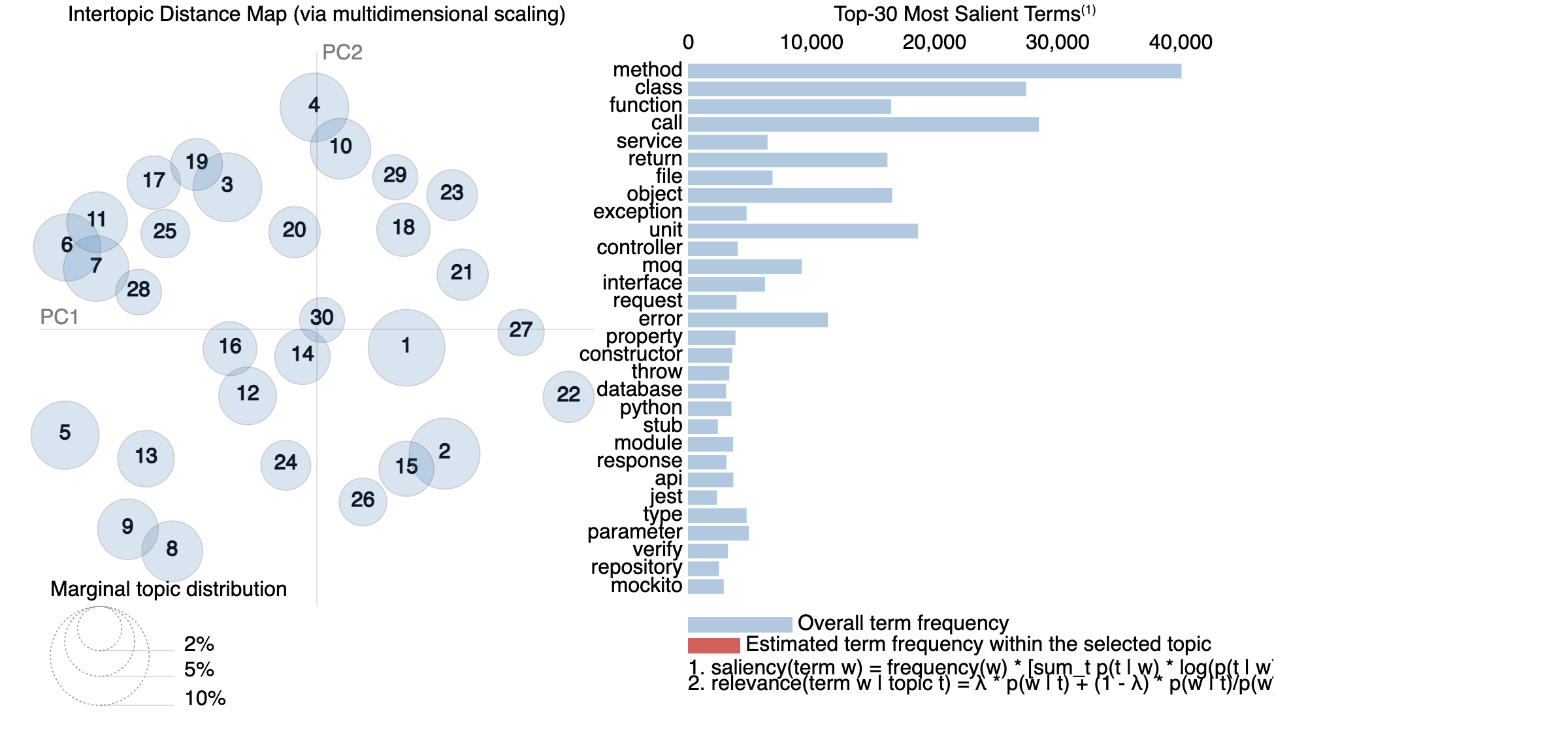}
        \caption{TTopic visualization for 30-topic LDA model using pyLDAvis.}
        \label{fig:30Topics}
    \end{subfigure}

    \caption{Topic modeling analysis including coherence graph and topic visualizations for different LDA models.}
    \label{fig:merged-lda-vertical}
\end{figure*}

In our research, we used a mixed-method approach as shown in Figure \ref{fig:method}.  We first develop a mocking tag set to identify relevant posts. In this context, a post refers to a \textsc{StackOverflow} thread, including the question (title and body) and its associated answers. 

The fourth and last authors collaborated to identify appropriate tags associated with mocking frameworks. Initially, we collected all \textsc{StackOverflow} posts tagged with the keyword \textit{mocking}. However, we noticed that many mocking-related questions were tagged with other tags in addition to \textit{mocking}. That is why we analyzed the co-occurring tags in the posts that were initially collected. Our goal was to construct a comprehensive tag set that would capture the majority of posts related to mocking.

To begin, we identified the key programming languages from the co-occurring tags that supports mocking frameworks, including \textit{Java}, \textit{Ruby}, \textit{Python}, \textit{JavaScript}, \textit{PHP}, \textit{C\#}, \textit{C++}, \textit{Go}, \textit{Swift}, \textit{Kotlin}, and \textit{Scala}. We then researched the popular mocking frameworks associated with these languages through documentation and relevant websites. This process led to the inclusion of 18 tags: \textit{mocking}, \textit{pytest-mock}, \textit{Mockito}, \textit{JMock}, \textit{EasyMock}, \textit{MockK}, \textit{Nock}, \textit{unittest.mock}, \textit{Moq}, \textit{NSubstitute}, \textit{FakeItEasy}, \textit{JustMock}, \textit{FakeIt}, \textit{Mockery}, \textit{GoMock}, \textit{Cuckoo}, \textit{Mockingbird}, and \textit{GoogleMock}. Any question tagged with at least one of these tags was classified as related to mocking and subsequently included in our dataset.

We used Stack Exchange Data Explorer (SEDE)\footnote{\url{https://data.stackexchange.com/stackoverflow/query/new} last accessed: 12-May-2025} to automate our data collection process. The dataset spans from August 7, 2008, to August 9, 2024. We obtained this corpus by querying all posts on Stack Overflow tagged with any of the tags in our tag set. We have collected both quantitative data (e.g., answer count, question view count, upvotes) and qualitative data (such as question titles, bodies, and answer bodies). Using the Stack Exchange API, we retrieved data for 25,302 questions, along with 35374 answers, and 40,207 comments.

These posts are then extracted and preprocessed by refining their titles and bodies for further analysis. Consequently, we applied the Latent Dirichlet Allocation (LDA) modeling technique~\cite{blei2003latent} to group the posts by dominant topics, followed by labeling and organizing these topics into a hierarchical structure.
\section{Approch, Analysis, and Results}
\label{sec:results}
This section outlines our approach for addressing each research question along with the corresponding findings.

\textbf{RQ1: What are the most common topics discussed about \textit{Mocking} on Stack Overflow?}

\textbf{Motivation:} Mocking frameworks introduce challenges that differ from those of general software development, as their effective use requires familiarity with the core testing principles, object-oriented programming concepts and specialized mocking techniques. We are therefore interested in understanding how these unique difficulties are reflected in developer discussions—particularly on Q\&A platforms like \textsc{StackOverflow}. In this research question, our aim is to uncover the key topics in mocking-related discussions, and identify emerging trends.

\textbf{Approach:} After constructing the \textit{Mocking} dataset, we filter irrelevant information to reduce noise in raw posts, which can interfere with model learning~\cite{barua2014developers}. We concatenate the question title and the question body of each \textsc{StackOverflow} post to construct a single, cohesive text string. In the data pre-processing step, we removed code snippets, HTML tags, and URLs, using regular expressions. Next, we clean up the text by removing email addresses, punctuations, and extra spaces. We applied the NLTK stopword corpus~\cite{loper2002nltk} to remove common words such as ‘a', ‘is', and ‘the' from the posts, aiming to reduce noise in the data, following methodologies from previous studies~\cite{haque2020challenges, zahedi2020mining, abdellatif2020challenges}. Subsequently, we constructed bigram models using Gensim~\cite{vrehuuvrek2022gensim}, as bigram modeling has been shown to improve text processing quality. We also applied lemmatization~\cite{honnibal2017spacy} to convert words to their base form (e.g., ‘development' to ‘develop'). This process results in a cleaned dataset that is then used in the next step for topic modeling.

In the next step, we applied LDA topic modeling to identify mocking-related topics. A key challenge in using LDA is determining the optimal number of topics (K). If K is too large, the topics may become too specific, making meaningful analysis difficult. In contrast, a small K value can result in overly broad topics that merge discussions from different areas. To generate topics, we applied the Latent Dirichlet Allocation (LDA) algorithm~\cite{blei2003latent} implemented in the MALLET\footnote{\url{https://people.cs.umass.edu/~mccallum/mallet/} last accessed: 12-May-2025} library~\cite{mccallum2002mallet}. MALLET has been used successfully in similar studies, producing reliable results~\cite{barua2014developers, rosen2016mobile}. Therefore, the topic coherence calculation method~\cite{mimno2011optimizing} was more effective in identifying an optimal topic number, as it generated more coherent topic distributions for our dataset. 

Consistent with previous research~\cite{barua2014developers, ahmed2018concurrency, rosen2016mobile, yang2016security}, we set the LDA hyperparameters to $\alpha = 50/K$ and run each model for 500, 1000, and 2,000 iterations. We implemented LDA with K values ranging from 1 to 100 in increments of 1 and computed topic coherence scores as shown in Figure~\ref{fig:coherence-graph}, where higher scores indicate better topic separation. The coherence metric assesses topic interpretability by using confirmation measures and is strongly correlated with human understanding~\cite{roder2015exploring}. Next, we recorded the corresponding coherence scores, finding that K values between 10 and 30 yield to similar scores. Therefore, we generated visualizations for the top 30 values for $K$ using LDAvis~\cite{sievert2014ldavis} to further analyze the topic structures. LDAvis provides an interactive representation of topic distributions, allowing us to explore the relationships and separability between topics. 

After extensive discussions, the first and second authors determined the optimal number of topics by closely analyzing the pyLDAvis visualizations for different values of $K$. We compared pyLDAvis visualizations specifically for topic models that exhibited high coherence scores. As shown in Figure~\ref{fig:15Topics} and Figure~\ref{fig:30Topics}, when comparing the visualizations for 15 and 30 topics, it became evident that the 30-topic model provided a more refined separation between topics. The intertopic distance map for 30 topics showed a greater degree of disjointness between topic circles, indicating a reduction in overlap and more clearly defined thematic boundaries. Furthermore, the term saliency and relevance graphs revealed a more granular and interpretable distribution of key terms between distinct topics in the 30-topic model.

Next, we used the open card sorting method~\cite{fincher2005making}, following previous studies~\cite{yang2016security, bagherzadeh2019going} to determine topic labels. The first and second authors initially labeled the topics independently. During the card-sorting phase, the authors iteratively refined these labels through collaborative discussions. Initially, we labeled 15 topics but found considerable noise and overlapping content. Later, we increased the number of topics and noted that the coherence scores remained relatively stable up to 30 topics. Furthermore, visual inspection using the pyLDAvis graph showed better topic separation at 30 topics. 

Based on these observations, we selected 30 topics for further analysis. We examined the top 10 words of each topic, as recommended by Agrawal et al.~\cite{agrawal2018wrong}, and analyzed 30 highly relevant posts (i.e., those with the highest probability of belonging to a particular topic) to determine an appropriate topic label. The process was repeated until mutual agreement was reached. More than 10 iterations were conducted, with discussions taking place through email exchanges and online communication tools such as Zoom. Once the topic labeling was finalized, we proceeded with a higher-level categorization. The topics were grouped on the basis of conceptual similarity.

\begin{figure*}[t] 
    \centering
    \includegraphics[width=\textwidth,keepaspectratio]{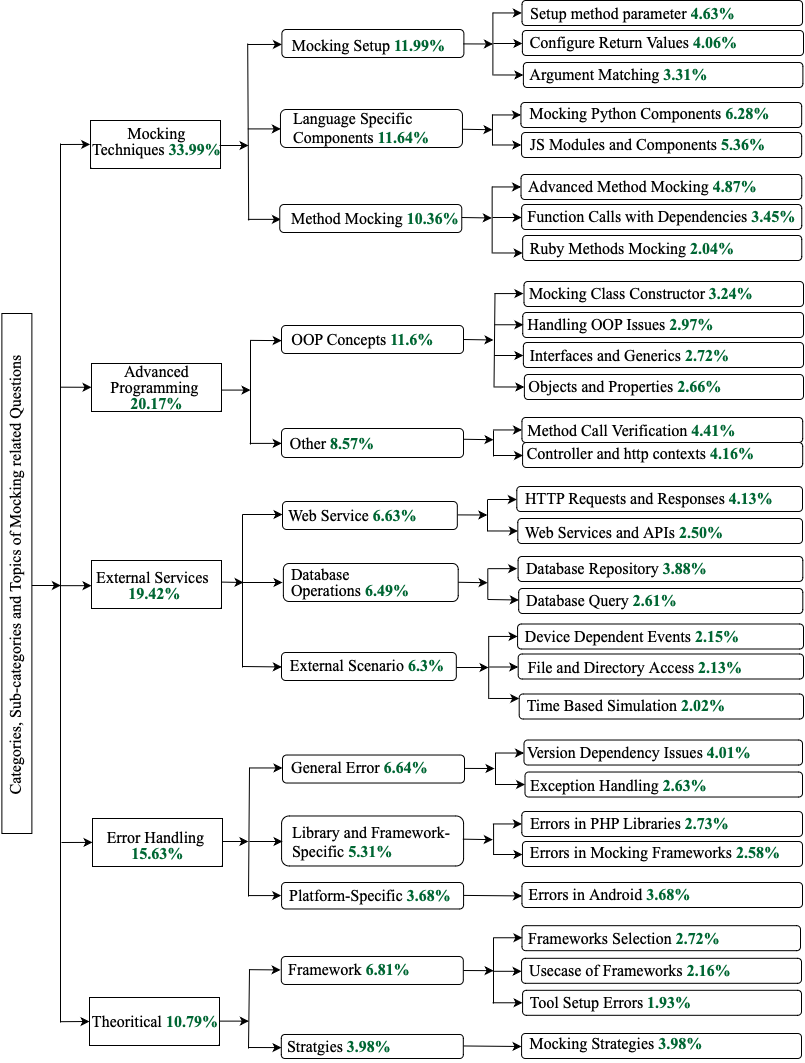}
    \caption{Hierarchy of Categories, Sub-Categories and Topics for Mocking related SO posts.}
    \label{fig:hierarchy}
\end{figure*}

\textbf{Results:}
After merging, we present the 5 major categories, 13 sub-categories and 30 topics for Mocking-related questions in Figure~\ref{fig:hierarchy}. The categories and sub-categories are ordered in descending order. Furthermore, Table~\ref{tab:mockito_topics} provides a detailed overview of 30 topics along with the top keywords and the number of posts. Next, we discuss the topics along with some example questions for each topic.
\subsection{Mocking Techniques}
Out of all topics, 33.99\% belong to this category, which focuses on setting up and handling mock objects in various programming languages. This category is divided into three sub-categories:  
\subsubsection{Mocking Setup (11.99\%)} This sub-category includes discussions on configuring mock objects and handling method parameters. Topics include:
\begin{itemize}
    \item \textbf{Setup Method Parameter} (4.63\%) discusses issues related to setting up method parameters, particularly challenges with mocking methods that have complex parameters. A common issue involves Moq throwing an error due to an invalid callback setup when parameters are passed as discussed in \href{https://stackoverflow.com/q/47356353}{Q47356353}\footnote{\url{https://stackoverflow.com/q/47356353} last accessed: 12-May-2025}.
    \item \textbf{Configure Return Values} (4.06\%) contains discussions on how to configure return values for functions or mocked methods to ensure expected outputs in various scenarios. For example, in \href{https://stackoverflow.com/q/36260465}{Q36260465}\footnote{\url{https://stackoverflow.com/q/36260465} last accessed: 12-May-2025} a notable issue is about configuring Moq to mock the \texttt{.Equals} method in C\#, which always returns true despite attempts to override it.
    \item \textbf{Argument Matching} (3.31\%) discusses about argument matching in unit testing, focusing on techniques to mock method calls, handle parameters and assert expected arguments. For example, \href{https://stackoverflow.com/q/25495912}{Q25495912} discusses the struggle to match arguments properly when verifying mock calls. 
\end{itemize}
\subsubsection{Language-Specific Components (11.64\%)} This sub-category focuses on mocking techniques in different programming languages. Topics include:
\begin{itemize}  
    \item \textbf{Mocking Python Components} (6.28\%) covers various techniques for mocking Python components such as functions, methods, static methods, decorators, and objects in unit tests. For example, \href{https://stackoverflow.com/q/31984449}{Q31984449} focus on mocking global functions used within Python classes. 
    \item \textbf{JS Modules and Components} (5.36\%) focuses on mocking dependencies in modern JavaScript applications. For example, \href{https://stackoverflow.com/q/62125544}{Q62125544} discusses a frequent issue in Jest failing to spy on functions due to undefined references.
\end{itemize}
\subsubsection{Method Mocking (10.36\%)} This sub-category covers discussions on advanced mocking techniques related to methods in different programming environments. Topics include:
\begin{itemize}
    \item \textbf{Advanced Method Mocking} (4.87\%) focuses on discussions on advanced techniques for mocking static, private, and external method calls. For example,\\     \href{https://stackoverflow.com/q/68274427}{Q68274427} discusses the problem of mocking methods within the same class during JUnit tests in Spring Boot applications.
    \item \textbf{Function Calls with Dependencies} (3.45\%) mentions testing of function calls with dependencies, focusing on mocking and verifying the behavior of functions with internal or external dependencies. For example,, \href{https://stackoverflow.com/questions/44922162/how-to-test-nested-mocked-functions}{Q44922162} discusses how to test nested mocked functions in Jest.
    \item \textbf{Ruby Methods Mocking} (2.04\%) contains discussions on mocking and stubbing methods in Rails applications using framework like RSpec, focusing on setting expectations, handling dependencies, and troubleshooting mocking issues. For example, \href{https://stackoverflow.com/q/1004832}{Q1004832} discusses the issue of mocking \texttt{request.subdomains} method, to facilitate functional testing.
\end{itemize}
\subsection{Advanced Programming} Advanced Programming (20.17\%) addresses issues related to object-oriented programming, interfaces, and constructors. This category consists of two sub-categories: 
\subsubsection{Object Oriented Programming Concepts} This sub-category describes different OOP concepts in the context of unit testing and mocking. Topics include: 
\begin{itemize} 
    \item \textbf{Mocking Class Constructor} (3.24\%) covers techniques for mocking class constructors in unit tests, with a focus on using frameworks to handle object creation and dependency injection effectively. For example, \href{https://stackoverflow.com/questions/60517635/mocking-generics-that-implement-multiple-interfaces}{Q60517635} discusses how to mock a parameter passed into a constructor using Mockito and PowerMock. 
    \item \textbf{Handling OOP Issues} (2.97\%) discusses various techniques for mocking abstract classes, base classes, and methods. For example, \href{https://stackoverflow.com/questions/6101030/how-to-create-mock-for-the-abstract-base-class-using-moq-framework}{Q6101030} discusses how to mock an abstract base class to skip logic in its constructor when testing a derived class.
    \item \textbf{Interfaces and Generics} (2.72\%) explores challenges and complexities in working with interfaces and generics in programming, specifically addressing issues in mocking. For example, \href{https://stackoverflow.com/questions/16153494/mocking-generics-that-implement-multiple-interfaces}{Q16153494} discusses the difficulty of mocking a generic class in C\# that implements multiple interfaces
    \item \textbf{Objects and Properties} (2.66\%) covers discussions of accessing and modifying private or protected members using reflection and mocking frameworks. For example, \href{https://stackoverflow.com/questions/14762362/how-to-pass-mock-object-to-another-mock-object-constructor}{Q14762362} focuses on how to pass a mock object to another mock object's constructor in unit tests.  
\end{itemize}
\subsubsection{Other} This sub-category covers the topic on controller testing and method call verification outlined as follows:
\begin{itemize}
    \item \textbf{Method Call Verification} (4.41\%) mentions about verifying whether specific method calls occur with expected arguments, order, and frequency in unit tests. For example, \href{https://stackoverflow.com/q/51691913}{Q51691913} discusses verifying the exact sequence of method calls using \texttt{Mockito.inOrder} and ensuring no additional calls occur by iterating over expected invocations and using\\ \texttt{verifyNoMoreInteractions()}.
    \item \textbf{Controller and http contexts} (4.16\%) discusses unit testing controllers and handling HTTP contexts using mocking frameworks. For example, The question \href{https://stackoverflow.com/q/50277705}{Q50277705} discusses unit testing ASP.NET Core controllers with session handling using mocks.  
\end{itemize}
\subsection{External Services} External Services (19.42\%) includes discussions on API interactions, HTTP requests, and database queries. This category is divided into three sub-categories:  
\subsubsection{Web Service} This sub-category discusses various methods for mocking web services, APIs, and HTTP requests or responses. Topics include:
     \begin{itemize} 
     \item \textbf{HTTP Requests and Responses} (4.13\%) covers various techniques for mocking HTTP requests and responses in different testing frameworks and languages, including Flask, FastAPI, EasyMock, and PowerMock, with a focus on ensuring correct behavior without making actual API calls. For example, \href{https://stackoverflow.com/questions/60940497/mocking-a-flask-post-request-with-json-arguments-using-pytest-and-pytest-mock}{Q60940497} focus on testing the input/output of a Flask POST request with JSON arguments using pytest and pytest-mock, and the challenge of mocking outcomes while ensuring proper function behavior without running the actual server.
     \item \textbf{Web Services and APIs} (2.50\%) discusses various methods for mocking web services, and APIs in applications, using tools like pytest, JMock, EasyMock, and gRPC, with a focus on mocking external calls, verifying parameters, and handling service layers in frameworks like Spring MVC. For example, \href{https://stackoverflow.com/questions/68365158/how-do-i-create-a-mock-context-for-grpc-unit-testing-python}{Q68365158}\footnote{\url{https://stackoverflow.com/questions/68365158/how-do-i-create-a-mock-context-for-grpc-unit-testing-python} last accessed: 12-May-2025} discusses how to create a mock context for gRPC unit testing in Python, as gRPC normally generates a context with requests.
\end{itemize} 
\subsubsection{Database Operations} This sub-category covers mocking techniques for database repositories and queries. Topics include:
    \begin{itemize} 
     \item \textbf{Database Repository} (3.88\%) discusses various challenges and solutions for unit testing database repositories using Moq, including mocking Entity Framework operations, handling CRUD operations, and managing relationships within models. For example, \href{https://stackoverflow.com/questions/28272400/unit-testing-add-operation-of-a-dbcontext-using-moq}{Q28272400} discusses about unit testing the Add operation of a dbContext using Moq in xUnit, specifically dealing with context being set to null after SaveChanges().
     \item \textbf{Database Query} (2.61\%) covers various approaches and tools for mocking and handling database queries across different databases and frameworks. For example, \href{https://stackoverflow.com/questions/73074795/how-to-see-sql-statement-in-a-test-that-im-using-mock}{Q73074795} discusses how to view the raw SQL generated by a query in a test where a mocked database connection is used.
  \end{itemize}
\subsubsection{External Scenario} This sub-category discusses mocking techniques for external dependencies such as file systems, device-dependent events, and time-based behaviors across various programming languages, as discussed below. 
\begin{itemize} 
        \item \textbf{Device Dependent Events} (2.15\%) covers various challenges and solutions related to mocking and unit testing in Android applications, such as, handling device-dependent events and testing Android components such as UI and EventBus. The \href{https://stackoverflow.com/questions/68563451/mock-or-avoid-cognito-authentication-and-group-permissions-for-pytest}{Q68563451} discusses how to mock or avoid Cognito authentication and group permissions while implementing pytest for a Flask app. 
        \item \textbf{File and Directory Access} (2.13\%) focuses on mocking file systems, file reads/writes, and handling file-related operations in unit tests across languages like Python, Go, and OCaml, using techniques such as in-memory file systems, temporary files, and libraries like vfsStream and os.walk. For example, \href{https://stackoverflow.com/questions/50970776/how-to-mock-images-and-other-file-formats-with-vfst}{Q50970776} discusses how to mock images and other file formats. Like PDFs of any size and resolution using vfsStream in PHP for testing purposes.
        \item \textbf{Time Based Simulation} (2.02\%) discusses techniques for simulating time-dependent behavior in tests, particularly when working with threads, sleep functions, and time-based tasks in Python. For example, \href{https://stackoverflow.com/questions/70650921/mock-async-task-of-django-q}{Q70650921} discusses how to mock Django-Q's async tasks during testing to avoid actual task queueing, focusing on simulating task behavior without triggering real task execution. 
     \end{itemize}
\subsection{Error Handling} Error Handling (15.63\%) encompasses discussions on tackling challenges such as exception handling, version dependency issues, and framework-specific bugs. This category consists of three sub-categories:
\subsubsection{General Error} This sub-category focuses on exception handling and version dependency issues as follows.
     \begin{itemize} 
         \item \textbf{Version Dependency Issues} (4.01\%) covers discussions on using mocking frameworks like Mockito, JMock, and EasyMock with testing tools such as JUnit, TestNG, and Spring, addressing issues related to annotations, dependencies, and version compatibility. For example, \href{https://stackoverflow.com/questions/1756468/using-spring-junit4-and-jmock-together}{Q1756468} focuses on the usage of JMock 2 with Spring in JUnit4 tests, seeking guidance on whether a Spring test runner supports JMock.
         \item \textbf{Exception Handling} (2.63\%) discusses the techniques and practices used in Java programming to handle runtime errors, including the use of try-catch blocks, throws, and custom exception classes. For example, \href{https://stackoverflow.com/questions/58263420/cant-mock-an-exception-with-pytest}{Q58263420} discusses the difficulty of testing a \texttt{GetoptError} exception using \texttt{pytest}.
         \end{itemize}
\subsubsection{Library and Framework-Specific} This sub-category mostly discusses library and framework specific errors. Topics include:
     \begin{itemize} 
     \item \textbf{Errors in PHP Libraries} (2.73\%) covers various issues and solutions related to mocking in PHP using Mockery and PHPUnit, specifically within Laravel applications. For example, \href{https://stackoverflow.com/questions/15407798/mockery-method-shouldrecieve-is-not-found-on-this-mock-object}{Q15407798} discusses an issue where the \texttt{shouldRecieve()} method is not found on a mock object while using the Mockery package with Laravel and PHPUnit.
     \item \textbf{Errors in Mocking Frameworks} ( 2.58\%) mentions various errors and issues related to mocking functions in different testing frameworks, including Google Mock, EasyMock, PowerMock, and Mockk, with a focus on resolving compilation errors and ambiguity. For example, \href{https://stackoverflow.com/questions/31167370/powermock-not-able-to-resolve-ambiguous-reference}{Q31167370} discusses an issue with
     \end{itemize} 
\begin{landscape}
\begin{table}[!htbp]
    \centering
    \caption{The Mocking Categories, Sub-categories, Topics Names, Top Ten Topic Words, and Number of Posts.}
    \label{tab:mockito_topics} 
    \small 
    \begin{tabular}{p{2.8cm} p{2.5cm} p{4cm} l c} 
        \toprule
        \textbf{Main Category} & \textbf{Sub-Categories} & \textbf{Topic Name} & \textbf{Topic Words} & \textbf{\# Posts} \\
        \midrule
        \multirow{3}{*}{\raggedright Mocking\ Techniques} 
         & \multirow{2}{*}{Mocking Setup} & \raggedright Configure Return Values & return, method, object, call, null, list, result, code, set, expect & 1026\\
         &  & \raggedright Setup method parameter & moq, method, setup, parameter, error, type, follow, return, expression, work & 1171\\
         &  & \raggedright Argument Matching & argument, pass, method, parameter, string, object, call, match, array, return & 837\\
         \cmidrule{2-5}
         & \multirow{2}{*} {Method Mocking} & Ruby Methods Mocking & stub, method, rspec, spec, object, code, call, work, return, block & 516\\
         &  & Advanced Method Mocking & method, class, call, static, private, unit, case, return, code, mockito & 1231\\
         &  & Function Calls with Dependencies & function, call, return, unit, write, code, inside, testing, make, case & 874\\
         \cmidrule{2-5}
         & \multirow{2}{*}{\shortstack[l]{Language Specific \\ Components}} & \raggedright Mocking Python Components & python, function, patch, module, work, import, code, call, follow, pyt & 1590\\
         &  & \raggedright JS Modules and Components & jest, component, module, function, work, error, file, call, follow, import & 1356\\
        \midrule
        \multirow{3}{*}{\raggedright Advanced Programming} 
         & \multirow{2}{*}{OOP Concepts} & Objects and Properties & object, property, set, class, create, variable, instance, attribute, field, access & 674\\
         &  & \raggedright Handling OOP Issues & class, method, call, base, abstract, create, instance, unit, virtual, code & 752\\
         &  & \raggedright Interfaces and Generics & interface, type, class, create, implement, implementation, method, object, moq, generic & 688\\
         &  & \raggedright Mocking Class Constructor & class, constructor, object, create, dependency, instance, inject, unit, pass, factory & 820\\
         \cmidrule{2-5}
         & \multirow{2}{*}{Other} & \raggedright Controller and http contexts & controller, unit, method, testing, action, moq, user, code, write, work & 1052\\
         &  & \raggedright Method Call Verification & call, method, verify, expect, fail, time, expectation, check, assert, object & 1117\\
        \midrule
        \multirow{3}{*}{\raggedright External Services} 
         & \multirow{2}{*}{Web Service} & \raggedright Web Services and APIs & service, client, server, call, unit, web, api, application, testing, create & 633\\
         &  & \raggedright HTTP Requests and Responses & request, response, api, call, http, server, nock, url, make, json & 1045\\
         \cmidrule{2-5}
         & \multirow{2}{*}{External Scenario} & \raggedright File and Directory Access & file, write, read, unit, code, create, datum, path, stream, content & 540\\
         &  & \raggedright Device Dependent Events & event, app, view, user, android, location, send, work, application, handler & 543\\
         &  & \raggedright Time Based Simulation & time, call, run, code, thread, task, unit, input, process, date & 510\\
         \cmidrule{2-5}
         & \multirow{2}{*}{Database Operations} & \raggedright Database Repository & repository, unit, method, moq, entity, service, context, work, code, add & 981\\
         &  & \raggedright Database Query & database, datum, unit, query, db, connection, create, object, write, testing & 661\\
        \midrule
        \multirow{3}{*}{\raggedright Error Handling} 
         & \multirow{2}{*}{General Error} & \raggedright Exception Handling & exception, throw, error, code, unit, method, line, follow, fail, null & 666\\
         &  & \raggedright Version Dependency Issues & class, mockito, junit, easymock, java, spring, error, work, code, follow & 1015\\
         \cmidrule{2-5}
         & \multirow{2}{*}{\shortstack[l]{Library and \\ Framework-Specific}} & \raggedright Errors in PHP Libraries & method, class, phpunit, error, mockery, laravel, call, model, work, follow & 691\\
         &  & \raggedright Errors in Mocking Frameworks & function, google, gmock, error, code, follow, compile, pointer, class, template & 652\\
         \cmidrule{2-5}
         & \multirow{2}{*}{\vspace{2mm}Platform-Specific}  & \raggedright Errors in Android  & service, client, server, call, unit, web, api, application, testing, create & 930\\
        \midrule
        \multirow{3}{*}{\raggedright Theoretical} 
         & \multirow{2}{*}{Framework} & \raggedright Usecase of Frameworks & moq, fake, nsubstitute, unit, code, work, fakeiteasy, framework, follow, nunit & 546\\
         &  & \raggedright Frameworks Selection & framework, testing, find, unit, question, understand, good, read, thing, answer & 689\\
         &  & \raggedright Tool Setup Errors & project, run, build, file, error, library, package, version, add, work & 489\\
         \cmidrule{2-5}
         & \multirow{2}{*}{\vspace{2mm}Stratgies} & \raggedright Mocking Strategies & code, unit, make, testing, change, write, question, approach, object, implementation & 1007\\
        \bottomrule
    \end{tabular}
\end{table}
\end{landscape} 
\begin{itemize}
\item[] PowerMock in Scala, where the user faces a compilation error due to an ambiguous reference while unit testing \texttt{SomeSystem.scala}.
\end{itemize}
\subsubsection{Platform-Specific} This sub-category focuses on errors relevant to Android platforms as below.
     \begin{itemize}    
     \item \textbf{Errors in Android} (3.68\%)  discusses various errors and issues encountered while running unit tests in Android and Kotlin, particularly related to mocking frameworks like MockK, Mockito. For example, \href{https://stackoverflow.com/questions/75559164/update-android-studio-electric-eel-the-mockk-spyk-is-failing}{Q75559164} discusses an issue where code works fine in Android Studio Dolphin but fails after updating to Android Studio Electric Eel due to compatibility issues with the mockk version and other dependencies. 
     \end{itemize}
\subsection{Theoretical} Theoretical (10.79\%) aspects cover discussions on framework selection, tool setup errors, and best practices in mocking.
    \subsubsection{Framework} This sub-category includes discussions on usecase, selection, and errors relevant to mocking frameworks. 
     \begin{itemize} 
     \item \textbf{Frameworks Selection} (2.72\%) focuses on various factors to consider when choosing a mocking framework, compares different mocking frameworks, and explores resources, pros, cons, and best practices for test-driven development (TDD). For example, \href{https://stackoverflow.com/questions/71113992/mocking-socket-activity-in-python}{Q71113992} discusses the user's struggle with using pytest-mock to test socket activity in Python, specifically aiming to check the number of calls to socket methods. 
     \item \textbf{Usecase of Frameworks} (2.16\%) discusses various use cases and comparisons of mocking frameworks like Moq, NSubstitute, FakeItEasy, and JustMock, including syntax differences, migration between frameworks. \href{{https://stackoverflow.com/questions/76109870/easymock-mock-new-instance-of-a-class-with-arguments}}{Q76109870} discusses transitioning from EasyMock to a different mocking framework (likely Moq or FakeItEasy), specifically focusing on how to mock a new instance of a class with arguments and troubleshoot issues encountered during the conversion process.
     \item \textbf{Tool Setup Errors} (1.93\%) covers common issues, failures, and errors encountered while unit testing Android applications, particularly with mocking frameworks like MockK, Mockito, and PowerMock, as well as problems related to test environment configurations. For example, \href{https://stackoverflow.com/questions/37169899/update-android-studio-electric-eel-mockk-spyk-failing}{Q37169899} discusses issues with installing Moq and MvvmCross.Tests packages in a Xamarin Studio project targeting a portable .NET framework, causing compatibility errors with the project's target profiles.
     \end{itemize}
\subsubsection{Strategies} This sub-category includes discussions on best practices for unit testing, mocking and dependency management as below.
     \begin{itemize} 
     \item \textbf{Mocking Strategies} (3.98\%) delivers best practices, opinions, and suggestions on how to effectively use mocking in unit testing, including structuring test code, reducing coupling, partial mocking concerns, dependency injection and when to use or avoid mocks. For example, \href{https://stackoverflow.com/questions/69557964/mocking-class-level-parameter-in-dummypy}{Q69557964} discusses usage of mocking to modify class-level parameters for testing purposes, where the user faces difficulties with mocking at the wrong level in their stripped-down example. 
    \end{itemize}
\begin{figure*}[t]
    \centering   \includegraphics[width=\textwidth,keepaspectratio]{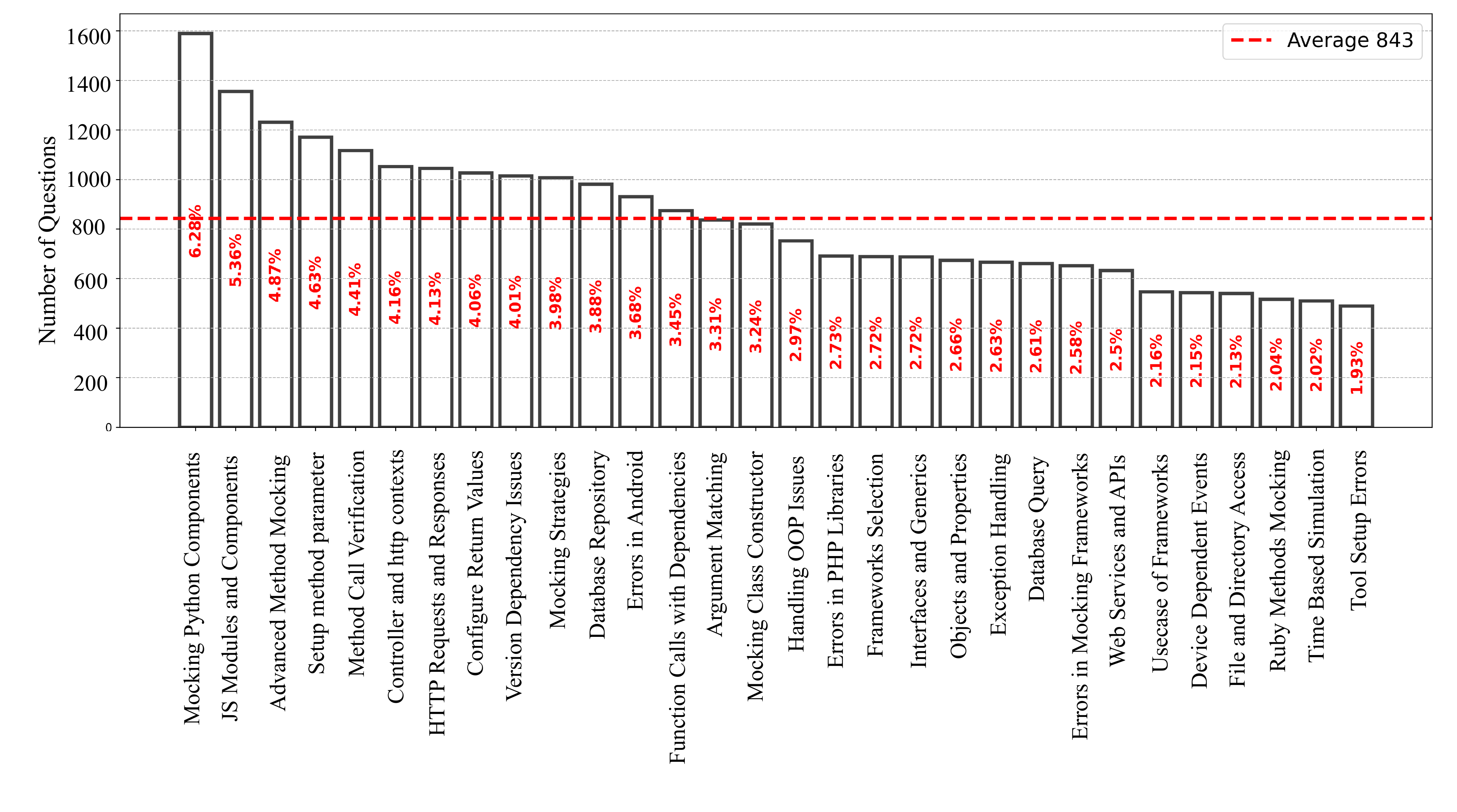}
        \caption{Distribution of Topics by Percentage of Questions}
    \label{fig:topics_distribution}
\end{figure*}

Next, we represent the distribution of topics by the percentage of questions asked in Figure~\ref{fig:topics_distribution}, with each topic represented by a bar. The most frequently asked topics include \textit{Mocking Python Components} (6.28\%), followed by \textit{JS Modules and Components} (5.36\%). Developers ask the most questions about \textit{Mocking Python Components}, highlighting common challenges such as patching functions, mocking modules, and handling imported components in unittest and pytest. Similarly, \textit{JS Modules and Components} focuses on mocking in JavaScript frameworks using tools like Jest, where developers frequently encounter issues with component mocking, import errors, and module resolution in environments like React, Angular, and Vue.

The high frequency of questions on both topics reflects the persistent complexities that developers face when working with mocking tools across different programming paradigms. However, the least frequent topics include \textit{Tool Setup Errors} (1.93\%) and \textit{Time-Based Simulation} (2.02\%) indicating fewer issues with setup and environment-related concerns. The red dashed line in Figure~\ref{fig:topics_distribution} marks the average questions (843), with \textit{ 13 of 30 topics exceeding this}, highlighting key challenges in method setup, verification, and API interactions.

\begin{tcolorbox}
\textbf{RQ1:} 
The analysis of mocking-related \textsc{StackOverflow} questions resulted in the identification of 5 major categories, 13 sub-categories, and 30 specific topics. The most prominent category, \textit{Mocking Techniques}, accounted for 33.99\% of the questions and includes topics such as parameter setup, language-specific components, and advanced method mocking. Other key categories include \textit{Advanced Programming}, \textit{External Services}, \textit{Error Handling}, and \textit{Theoretical} discussions. Each category captures diverse aspects of mocking, such as handling constructors, API testing, database interactions, framework-specific errors, and conceptual understanding of mocking strategies.
\end{tcolorbox}

\textbf{RQ2: How have \textit{Mocking}-related discussions on \textsc{StackOverflow} evolved over the years?}\\
\textbf{Motivation:} Inspired by previous research \cite{uddin2021empirical, abdellatif2020challenges, peruma2022refactor}, we now investigate the evolution of various categories and topics in mocking-related questions. This examination enables us to trace the topics that currently demand more focus. By identifying these emerging trends, educators can tailor training materials more effectively and practitioners can stay updated with relevant and timely content.

\textbf{Approach:} We have analyzed the trends and temporal effects of different mocking categories and topics. We evaluated the annual probabilities for each category and topic separately. Additionally, we computed the relative probability of them compared to the others. The probability of a category \( c \) in a given year \( y_i \) is calculated using Equation \ref{eqn:probability}, which considers the number of questions related to that category during the data collection period (2008 to 2024). In this equation, \( N_{c y_i} \) represents the number of questions associated with category \( c \) for year \( y_i \).

The graph derived from Equation \ref{eqn:probability}\textemdash{}the Probability Distribution Function (PDF)\textemdash{}illustrates the yearly trends of various categories on \textsc{Stackoverflow}. However, PDFs are typically non-monotonic and exhibit a zigzag pattern, which can complicate interpretation. To improve clarity, we employ the Cumulative Distribution Function (CDF), as described in Equation \ref{eqn:cdf}. In this case, \( C_{c y} \) is the cumulative sum of the probabilities of a category from 2008 to year \( y \). Similarly, the relative probability of a category is calculated as shown in Equation \ref{eqn:relative_prob}, where \( N_{c y_i} \) and \( N_{C y_i} \) represent the number of questions posted for category \( c \) and all Categories in year \( y_i \), respectively. It is important to note that a CDF cannot be generated for this equation, as the CDF is only applicable when evaluating a single category independently. Consequently, we have used Equation \ref{eqn:cdf} and \ref{eqn:relative_prob} to compute the annual and relative probabilities of each topic.
\begin{equation}
P_{c y_i} = \frac{N_{c y_i}}{\sum_{y_i=2008}^{2024} N_{c y_i}} \label{eqn:probability}
\end{equation}
\begin{equation}
C_{c y} = \sum_{y_i=2008}^{y} P_{c y_i} \label{eqn:cdf}
\end{equation}
\begin{equation}
A_{c y_i} = \frac{N_{c y_i}}{N_{C y_i}} \label{eqn:relative_prob}
\end{equation}
\begin{figure*}
    \centering
    \begin{subfigure}{0.48\textwidth}
        \centering
        \includegraphics[width=\linewidth,keepaspectratio]{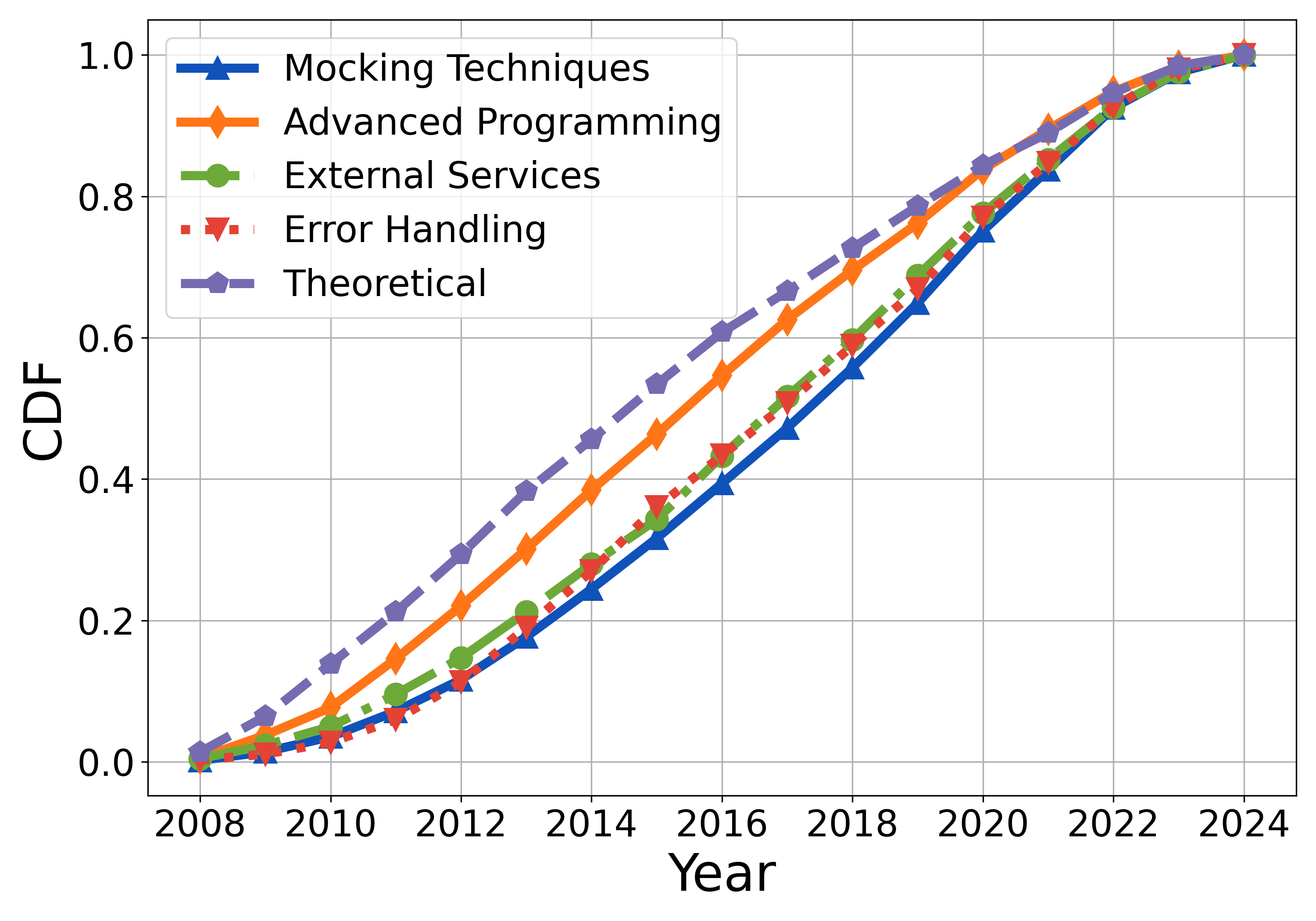}
        \caption{}
        \label{fig:cdf_categories}
    \end{subfigure}
    \hfill
    \begin{subfigure}{0.48\textwidth}
        \centering
        \includegraphics[width=\linewidth,keepaspectratio]{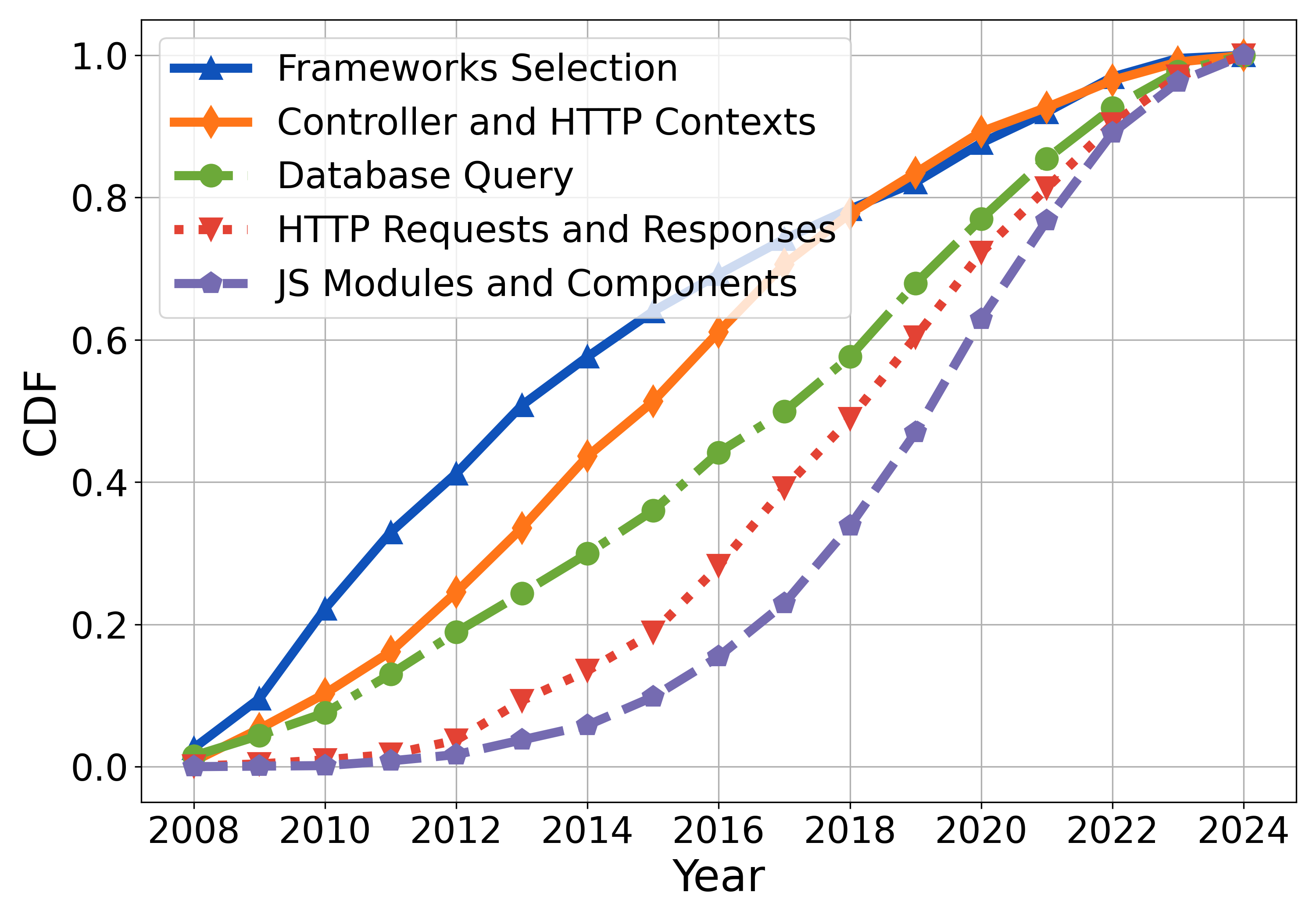}
        \caption{}
        \label{fig:cdf_topics}
    \end{subfigure}
    \caption{(a) Cumulative Distribution Function (CDF) of questions posted in different categories over the years. The \textit{Theoretical} and \textit{Advanced Programming} categories showed an early increase in submissions, while other categories followed a more gradual trend. (b) Cumulative Distribution Function (CDF) of questions posted in different mocking-related topics over the years. The Topics \textit{Frameworks Selection}, and \textit{Controller and HTTP Contexts} showed an early increase in submissions, while other categories followed a more gradual trend. To improve the clarity of the graph, only five chosen topics are displayed, with the remaining topics covered in the accompanying text.}
\end{figure*}
\begin{figure*}
    \centering
    \begin{subfigure}{0.48\textwidth}
        \centering
        \includegraphics[width=\linewidth,keepaspectratio]{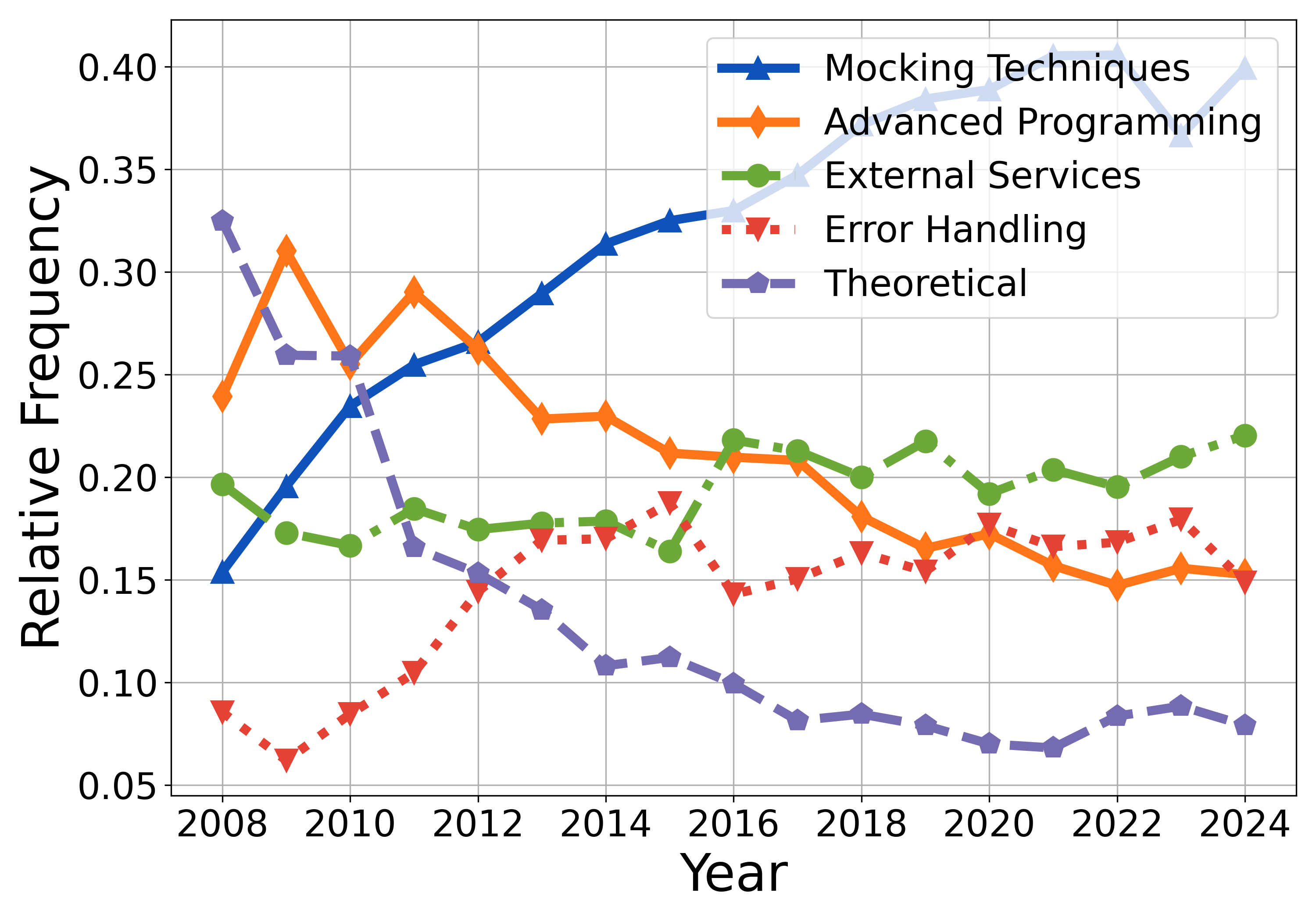}
        \caption{}
        \label{fig:relative_impact_category}
    \end{subfigure}
    \hfill
    \begin{subfigure}{0.48\textwidth}
        \centering
        \includegraphics[width=\linewidth,keepaspectratio]{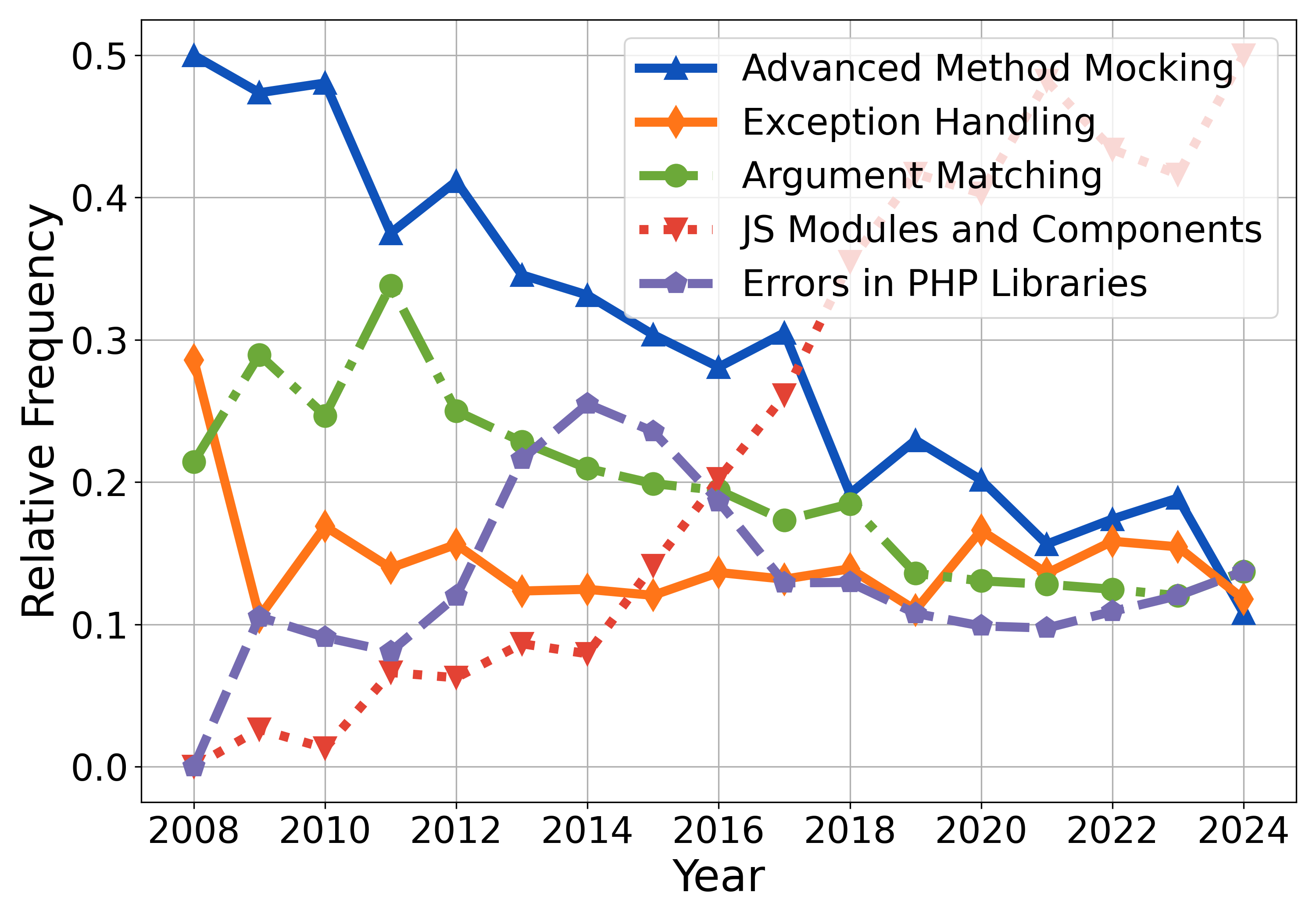}
        \caption{}
        \label{fig:relative_impact_topics}
    \end{subfigure}
    \caption{(a) Relative probability of question categories over time. The \textit{Mocking Techniques} category showed an increased impact by significantly rising, while the \textit{Theoretical} category had steadily declined in relevance since 2010.
    (b) Relative probability of different mocking-related topics over time. The \textit{JS Modules and Components} topic became increasingly dominant after 2014. To improve the clarity of the graph, only five chosen topics are displayed, with the remaining topics covered in the accompanying text}
    \label{fig:relative_mocking_combined}
\end{figure*}
\textbf{Results:} We present the cumulative distribution function (CDF) for five mocking categories over the years in Figure~\ref{fig:cdf_categories}. Among the five categories analyzed, the \textit{Theoretical} and \textit{Advanced Programming} category exhibit an early surge in submitted questions, standing apart from the other categories. By 2016, approximately 60\% of questions in \textit{Theoretical} and 50\% of the category \textit{Advanced Programming} had already been posted, indicating early interest. In contrast, categories like \textit{External Services}, \textit{Error Handling} and \textit{Mocking Techniques} saw a more gradual accumulation of questions, with a steady rise over time.

Next, we present the cumulative distribution function (CDF) of the posted questions for five selected topics over the years in Figure \ref{fig:cdf_topics}. Among these, \textit{Frameworks Selection} and \textit{Controller and HTTP Contexts} show early submission of questions, with almost 60\% and 50\% of their respective questions submitted by 2015. After 2018, the growth in new questions for these topics slows significantly with only 20\% of question submissions, indicating their maturity in the developer community. A similar trend was observed on topics such as \textit{Ruby Method Mocking}, \textit{Objects and Properties}, \textit{Argument Matching}, \textit{Interface and Generics}, \textit{Exception Handling}, etc. In contrast, \textit{HTTP Requests and Responses}, \textit{JS Modules and Components} and other topics (i.e., \textit{Mocking Python Components}, \textit{Tool Setup Errors}, \textit{File and Directory Access}, \textit{Function Call with Dependencies}) exhibit delayed but rapid growth, with a significant portion of questions posted after 2018, reflecting the rise of API-centric and modular front-end development practices. \textit{Database Query} shows a more uniform distribution across the entire timeline, similar to topics such as \textit{Configure Return Values}, \textit{Database Repository}, \textit{Version Dependency Issues}, etc. suggesting that the challenges in these topics have remained consistently relevant throughout the study period.

Although this provides insights into how different categories and topics evolved over time, it does not yet explain their relative importance compared to others. To address this, we analyzed the annual impact of each category. Figure~\ref{fig:relative_impact_category} illustrates the probability of categories over time, highlighting the dominance of \textit{Mocking Techniques}, which steadily increased to surpass other areas by 2014 and accounted for nearly 40\% of the total questions by 2024. In contrast, the \textit{Theoretical} category was initially prominent but declined after 2010, reflecting a shift toward practical concerns. \textit{Advanced Programming} and \textit{External Services} maintained stable but fluctuating impacts, while \textit{Error Handling} grew notably from 2010, emphasizing debugging and exception handling. Despite the presence of other categories, none has overtaken the sustained dominance of \textit{Mocking Techniques} in recent years.

Furthermore, we compared the relative probability of each topic over time in Figure \ref{fig:relative_impact_topics}. From 2008 to around 2015, \textit{Advanced Method Mocking} was the most dominant topic, consistently maintaining the highest relative frequency among all topics. However, it experienced a steady decline after 2015 and was significantly reduced by 2024. A similar trend was observed on topics such as \textit{Database Repository}, \textit{Database Query}, \textit{Method Call Verification}, \textit{Interfaces and Generics}, etc.  Meanwhile, \textit{JS Modules and Components} and other topics (i.e., \textit{Function Calls with Dependencies}, \textit{HTTP Requests and Response} and \textit{Errors in Android}) saw a dramatic rise beginning in 2014, eventually overtaking all other topics by 2024. \textit{Argument Matching} maintained moderate popularity in the early 2010s, peaking around 2011, but gradually declined afterward, similar to \textit{Web Services and APIs}, \textit{Device Dependent Events}, \textit{Ruby method Mocking}, etc. \textit{Errors in PHP Libraries} saw a sharp increase between 2011 and 2014 before falling back to lower levels in later years, a trend also observed in topics such as \textit{File and Directory Access}, \textit{Errors in Mocking Frameworks}, and \textit{Tool Setup Errors}. \textit{Exception Handling} remained relatively stable throughout the period, with small declines in the early stages, similar to \textit{Configure Return Values}, \textit{Handling OOP Issues}, and \textit{Controller and HTTP Contexts}.
\begin{tcolorbox}
\textbf{RQ2.} The categories consistently dominating mocking-related discussions are \textit{Mocking Techniques}, \textit{Advanced Programming}, and \textit{External Services}. In recent years, topics such as \textit{HTTP Requests and Responses}, \textit{JS Modules and Components}, and \textit{Mocking Python Components} have begun to gain prominence. Although \textit{Theoretical} discussions were initially prominent, their relevance declined significantly after 2010. Topics like \textit{Frameworks Selection} and \textit{Controller and HTTP Contexts} matured early and saw reduced activity after 2018.  
\end{tcolorbox}
\textbf{RQ3: How do \textit{Mocking}-related topics vary in terms of popularity and difficulty?}\\
\textbf{Motivation}: We explore whether \textit{Mocking} related topics exhibit unique trends in terms of popularity and difficulty. Specifically, we investigate how frequently these topics are discussed, how engaging they are with the community, and whether certain topics pose more challenges than others. Next, we assess the difficulty of each mocking-related topic based on how challenging it is for developers to receive satisfactory solutions. Understanding whether certain topics are more difficult to address will help identify areas that require more attention from the community. 

\textbf{Approach}: We evaluated the popularity of these questions by comparing them with a randomly selected sample of one million \textsc{StackOverflow} questions. Following the approach of Pinto et al. \cite{pinto2014mining}, we measured popularity using four key metrics: question score, number of answers, number of comments, and number of views. Although Pinto et al. originally included the \textit{favourites} metric, it is no longer supported on \textsc{StackOverflow}. The score metric reflects community feedback, where users can upvote or downvote a question $-$ each upvote adds $+1$, while each downvote subtracts $-1$ from the total score.

To enable a meaningful comparison with other \textsc{StackOverflow} questions, we assessed the relative popularity of mocking-related questions using the following formula:
\begin{equation}
    P = \frac{S + A + C + V}{4}
    \label{eq:relative_popularity}
\end{equation}
where $S$ is the relative score, $A$ is the relative number of answers, $C$ is the relative number of comments, and $V$ is the relative number of views. These relative values are calculated as:
\begin{equation}
    \text{relativeValue} = \frac{\text{avgMockito}}{\text{avgSO}}
    \label{eq:relative_factor}
\end{equation}
where $\text{avgMockito}$ represents the average value of a given factor (e.g., score) in the ‘Mocking' question set, and $\text{avgSO}$ represents the average of the same metric across non-mocking question set. For instance, if mocking-related questions have an average of 100 views and non-mocking questions have average 200 views, the relative view count would be 100/200 = 0.5. We computed this ratio for each of the four selected metrics. Then, using Equation \ref{eq:relative_popularity}, we combine them to derive an overall measure of relative popularity. A relative popularity score of 0.5 suggests that mocking-related questions are only half as popular as the average \textsc{StackOverflow} questions, whereas a score of 2 would indicate they are twice as popular.

Using a similar approach, we compute the relative popularity of answers for the mocking-related questions. Since answers on \textsc{StackOverflow} do not have specific metrics like view count or answer count, we adapted the equation based on the methodology proposed by Pinto et al. \cite{pinto2014mining}, as follows:
\begin{equation}
    P = \frac{S + C}{2}
    \label{eq:answer_popularity}
\end{equation}
To measure the difficulty level of each mocking-related topic, we adopt a two-metric approach widely used in previous research \cite{abdellatif2020challenges, haque2020challenges, uddin2021empirical}. The first metric is the percentage of posts without accepted answers (\% w/o accepted answers). On \textsc{StackOverflow}, users can mark a single answer as `accepted' if it sufficiently resolves their question. Therefore, we compute the percentage of posts that do not have an accepted answer for each topic. A higher percentage suggests that users find it harder to receive a satisfactory solution, indicating a higher difficulty level for that topic.

The second metric is the median time in hours required for an answer to be accepted, referred to as Median Time (Hrs.). This metric captures how long it typically takes for a post to receive an accepted answer, based on the time the answer is posted rather than when it is marked as accepted. A longer median time implies that it takes more effort or expertise to provide a satisfactory response, reflecting greater difficulty. To provide further context for these metrics, we categorize each question into one of three success statuses. A question is labeled as \textit{successful} if it has an accepted answer, \textit{ordinary} if it has at least one answer but none are accepted, and \textit{unsuccessful} if it remains completely unanswered. 

\textbf{Results}: Table~\ref{tab:results} visualizes the variation in question popularity and answer popularity across different topics. The metrics include relative question score, answer count, comment count, view count, and popularity for questions, as well as Score, Comment Count, and Popularity for answers. Mocking-related questions on \textsc{StackOverflow} are notably more popular than the average SO questions, with a normalized popularity score of \textit{1.17}. These questions receive higher scores \textit{1.63} and more views \textit{1.21} than typical posts, although the number of answers and comments is only slightly below average. This discrepancy\textemdash{}high engagement through scores and views, but relatively fewer answers and comments\textemdash{}suggests that while these questions are in high demand, they are more challenging to answer compared to other posts. Mocking answers also show strong performance, with high scores (1.47) and overall popularity (1.24).

These findings suggest that mocking-related content receives substantial attention and engagement, reflecting its relevance and importance within the developer community.
Notably, \textit{Framework Selection} exhibits the highest question score and popularity, suggesting significant community interest. Similarly, \textit{Method Call Verification} and \textit{Setup Method Parameter} show the highest answer scores, indicating these topics receive high-quality answers. However, despite high engagement, some topics, such as \textit{Database Query}, have relatively low answer scores, suggesting potential difficulty in resolving these issues.

Table~\ref{tab:topic_difficulty} presents the difficulty measures for answering mocking-related \textsc{StackOverflow} questions, highlighting the percentage of questions without accepted answers and the median time it takes to receive an accepted answer. Topics such as \textit{HTTP Requests and Responses} (58.85\%), \textit{JS Modules and Components} (58.7\%), and \textit{Web Services and APIs} (55.77\%) exhibit the highest percentage of questions without accepted answer. Correspondingly, these topics also record the longest median response times\textemdash{}7 hours, 6 hours, and 5 hours, respectively. In addition, as shown in Table~\ref{tab:mocking_analysis}, these topics demonstrate notably high percentages of unsuccessful questions—18.56\%, 18.64\%, and 19.4\%\textemdash{}further suggesting that these topics are difficult to resolve. 

Again, \textit{Errors in Android} have 50.86\% of questions without accepted answers and the highest (20.65\%) percentage of unsuccessful questions, as shown in Table~\ref{tab:mocking_analysis}, indicating that Android platform-related issues are also largely unresolved. In contrast, as illustrated in Table~\ref{tab:mocking_analysis}, topics such as \textit{Frameworks Selection} (66.62\%) and \textit{Use Case of Frameworks} (66.3\%) demonstrate relatively high success rates, indicating that questions in these areas are more likely to receive effective solutions. Whereas, topics such as \textit{Tool Setup Errors} (45.81\%) have lower success rates, highlighting potential difficulties in resolving issues in these domains. 

\textbf{Comparison with other SE Fields:} 
\textsc{StackOverflow} posts have been extensively studied across various domains. Each study focuses on analyzing discussions within a specific domain, revealing variations in question distribution and engagement levels among developers. To systematically compare different domains, we examine studies that analyze SO posts, identifying six key metrics. We compare three metrics that assess the popularity of topics: (1) total number of posts analyzed, (2) average views, and (3) average scores. As the metric Favorites is not available in \textsc{StackOverflow} now, we did not compare this metric. We also compare two additional metrics to measure topic difficulty: (5) the percentage of unanswered questions and (6) the median time to accept an answer. Instead of replicating prior analyses, we relied on metric values as reported in existing studies to perform our comparison across domains.

Among the available literature, five studies provide the five metrics, covering big data~\cite{bagherzadeh2019going}, chatbots~\cite{abdellatif2020challenges}, security~\cite{yang2016security}, mobile apps~\cite{rosen2016mobile}, and concurrency~\cite{ahmed2018concurrency}. Although blockchain~\cite{wan2019programmers} and deep learning~\cite{han2020programmers} studies also analyze SO posts, they do not report all six metrics. Table~\ref{tab:fields_stats} compares the popularity and difficulty of various Software Engineering (SE) fields. In terms of popularity, Mocking has 25,302 posts with an average view count of 4,696.46, placing it higher than IoT (1,320.3 views), Security (2,461.1 views) and other fields. This high view count indicates strong community interest, suggesting Mocking is a commonly referenced and relevant topic. 

It also has a relatively high average score of 4.42, outperforming Chatbot (0.7), IoT (0.8), and Big Data (1.4), which reflects the usefulness and clarity of questions in this domain. On the difficulty side, Mocking has 45.41\% of posts without accepted answers, indicating a moderate level of difficulty. While this is lower than fields like Chatbot (67.7\%), IoT (64\%), and Big Data (60.3\%), it still shows some challenge. Additionally, it takes a median of 1.28 hours to get an accepted answer, which is faster than IoT (2.9 hours), Big Data (3.3 hours), and Chatbot (14.8 hours). Overall, Mocking stands out for its relatively high popularity and lower difficulty compared to some other fields in SE.
\begin{landscape}
\begin{table}[!htbp]
    \centering
    \caption{Relative Popularity and Popularity Factors of Mocking Questions and Answers}
    \label{tab:results}
    \normalsize
    \begin{tabular}{p{5.5cm}p{1.5cm}p{2cm}p{2cm}p{2cm} p{2cm}p{1.5cm}p{2cm}p{1.5cm}}
        \toprule
        \textbf{Topic} & \textbf{$Q_{Score}$} & \textbf{$Q_{AnswerCount}$} & \textbf{$Q_{CommentCount}$} & \textbf{$Q_{ViewCount}$} & \textbf{$Q_{Popularity}$} & \textbf{$A_{Score}$} & \textbf{$A_{CommentCount}$} & \textbf{$A_{Popularity}$} \\
        \midrule
        Configure Return Values & 1.88 & 0.92 & 0.91 & 0.20 & 0.98 & 1.28 & 1.09 & 1.19 \\
        Setup Method Parameter & 2.93 & 0.96 & 0.92 & 0.29 & 1.27 & 2.31 & 1.11 & 1.71 \\
        Argument Matching & 2.21 & 0.97 & 0.71 & 0.22 & 1.03 & 1.79 & 1.08 & 1.44 \\
        Ruby Methods Mocking & 2.31 & 1.04 & 0.53 & 0.31 & 1.05 & 1.67 & 1.10 & 1.39 \\
        Advanced Method Mocking & 2.50 & 1.05 & 0.88 & 0.16 & 1.15 & 1.88 & 1.20 & 1.54 \\
        Function Calls with Dependencies & 1.25 & 0.80 & 0.79 & 0.13 & 0.74 & 0.90 & 1.02 & 0.96 \\
        Mocking Python Components & 2.16 & 0.82 & 0.73 & 0.15 & 0.97 & 1.59 & 1.05 & 1.32 \\
        JS Modules and Components & 1.94 & 0.82 & 0.55 & 0.12 & 0.86 & 1.26 & 1.00 & 1.13 \\
        Objects and Properties & 1.74 & 0.99 & 0.88 & 0.27 & 0.97 & 1.22 & 1.14 & 1.18 \\
        Handling OOP Issues & 1.61 & 0.99 & 0.96 & 0.17 & 0.93 & 1.02 & 1.13 & 1.08 \\
        Interfaces and Generics & 2.3 & 1.00 & 0.88 & 0.33 & 1.13 & 1.42 & 1.21 & 1.32 \\
        Mocking Class Constructor & 1.63 & 1.04 & 0.92 & 0.22 & 0.95 & 1.26 & 1.17 & 1.21 \\
        Controller and http contexts & 1.84 & 0.96 & 0.82 & 0.22 & 0.96 & 1.68 & 1.25 & 1.47 \\
        Method Call Verification & 2.94 & 0.98 & 0.72 & 0.31 & 1.24 & \textbf{2.35} & 1.17 & \textbf{1.76} \\
        WebServices and APIs & 1.19 & 0.87 & 0.66 & 0.18 & 0.73 & 0.81 & 0.91 & 0.86 \\
        HTTP Requests and Responses & 1.45 & 0.83 & 0.58 & 0.12 & 0.74 & 0.96 & 0.88 & 0.92 \\
        File and Directory Access & 1.3 & 0.91 & 0.72 & 0.21 & 0.78 & 0.91 & 1.13 & 1.02 \\
        Device Dependent Events & 1.52 & 0.93 & 0.60 & 0.13 & 0.79 & 0.84 & 1.08 & 0.96 \\
        Time Based Simulation & 2.36 & 1.01 & 0.88 & 0.21 & 1.11 & 1.90 & 1.17 & 1.53 \\
        Database Repository & 1.25 & 0.93 & \textbf{1.01} & 0.22 & 0.86 & 0.93 & \textbf{1.35} & 1.14 \\
        Database Query & 1.12 & 0.93 & 0.79 & 0.14 & 0.75 & 0.76 & 1.02 & 0.89 \\
        Exception Handling & 2.14 & 0.95 & 0.94 & 0.23 & 1.06 & 1.94 & 1.17 & 1.55 \\
        Version Dependency Issues & 1.88 & 0.99 & 0.73 & 0.18 & 0.94 & 1.49 & 1.21 & 1.35 \\
        Errors in PHP Libraries & 1.06 & 0.80 & 0.68 & 0.13 & 0.67 & 0.87 & 1.02 & 0.94 \\
        Errors in Mocking Frameworks & 1.13 & 0.82 & 0.98 & 0.16 & 0.77 & 0.87 & 1.06 & 0.97 \\
        Errors in Android & 1.26 & 0.80 & 0.90 & 0.20 & 0.79 & 0.99 & 0.99 & 0.99 \\
        Usecase of Frameworks & 1.79 & 1.02 & 0.80 & 0.21 & 0.96 & 1.40 & 1.11 & 1.26 \\
        Framework Selection & \textbf{4.05} & \textbf{1.38} & 0.69 & \textbf{0.34} & \textbf{1.62} & 1.86 & 1.10 & 1.48 \\
        Tool Setup Errors & 1.44 & 0.87 & 0.79 & 0.15 & 0.81 & 1.15 & 1.06 & 1.11 \\
        Mocking Strategies & 2.31 & 1.23 & 0.88 & 0.38 & 1.20 & 1.01 & 1.29 & 1.15 \\
        \bottomrule
        All Mocking  Questions and Answers & 1.63 & 0.92 & 0.91 & 1.21 & 1.17 & 1.47 & 1.00 & 1.24 \\
        \bottomrule
    \end{tabular}
 \end{table}
\end{landscape}
\begin{table*}[t]
    \centering
    \caption{Statistics on Topics Without Accepted Answers and Median Accepted Time}
    \normalsize
    \begin{tabular}{l c c}
        \toprule
        \textbf{Topic} & \textbf{\% w/o Acc Ans} & \textbf{Median Time (Hrs.)} \\
        \midrule
        Configure Return Values & 36.94 & 1 \\
        Setup Method Parameter & 34.67 & 1 \\
        Argument Matching & 38.23 & 2 \\
        Ruby Methods Mocking & 39.73 & 3 \\
        Advanced Method Mocking & 46.3 & 1 \\
        Function Calls with Dependencies & 52.63 & 2 \\
        Mocking Python Components & 46.54 & 2 \\
        JS Modules and Components & 58.7 & \textbf{7} \\
        Objects and Properties & 41.39 & 1 \\
        Handling OOP Issues & 46.01 & 1 \\
        Interfaces and Generics & 41.42 & 1 \\
        Mocking Class Constructor & 42.07 & 1 \\
        Controller and HTTP Contexts & 40.97 & 1 \\
        Method Call Verification & 35.99 & 1 \\
        Web Services and APIs & 55.77 & 5 \\
        HTTP Requests and Responses & \textbf{58.85} & 6 \\
        File and Directory Access & 45.56 & 1 \\
        Device Dependent Events & 50.28 & 3 \\
        Time Based Simulation & 46.47 & 1 \\
        Database Repository & 41.08 & 1 \\
        Database Query & 53.25 & 4 \\
        Exception Handling & 47.22 & 1 \\
        Version Dependency Issues & 52.41 & 2 \\
        Errors in PHP Libraries & 48.63 & 3 \\
        Errors in Mocking Frameworks & 48.16 & 3 \\
        Errors in Android & 50.86 & 3 \\
        Use Case of Frameworks & 33.7 & 2 \\
        Frameworks Selection & 33.38 & 1 \\
        Tool Setup Errors & 54.19 & 4 \\
        Mocking Strategies & 39.52 & 1 \\
        \bottomrule
    \end{tabular}
    \label{tab:topic_difficulty}
\end{table*}
\begin{table*}[t]
    \centering
    \caption{Summary of Topics and Question Success Rates}
    \label{tab:mocking_analysis}
    \small
    \begin{tabular}{lllll}
        \toprule
        \textbf{Topic} & \textbf{Total Questions} & \textbf{Successful \%} & \textbf{Ordinary \%} & \textbf{Unsuccessful \%} \\
        \midrule
        Configure Return Values & 1026 & 63.06 & 25.93 & 10.92 \\
        Setup Method Parameter & 1171 & 65.33 & 25.62 & 9.05 \\
        Argument Matching & 837 & 61.77 & 27.96 & 10.27 \\
        Ruby Methods Mocking & 516 & 60.27 & 31.01 & 8.72 \\
        Advanced Method Mocking & 1231 & 53.7 & 35.74 & 10.56 \\
        Function Calls with Dependencies & 874 & 47.37 & 35.58 & 17.05 \\
        Mocking Python Components & 1590 & 53.46 & 28.74 & 17.8 \\
        JS Modules and Components & 1356 & 41.3 & 39.31 & 19.4 \\
        Objects and Properties & 674 & 58.61 & 29.67 & 11.72 \\
        Handling OOP Issues & 752 & 53.99 & 34.84 & 11.17 \\
        Interfaces and Generics & 688 & 58.58 & 30.96 & 10.47 \\
        Mocking Class Constructor & 820 & 57.93 & 31.34 & 10.73 \\
        Controller and HTTP Contexts & 1052 & 59.03 & 28.9 & 12.07 \\
        Method Call Verification & 1117 & 64.01 & 27.22 & 8.77 \\
        Web Services and APIs & 633 & 44.23 & 37.12 & 18.64 \\
        HTTP Requests and Responses & 1045 & 41.15 & 40.29 & 18.56 \\
        File and Directory Access & 540 & 54.44 & 30.37 & 15.19 \\
        Device Dependent Events & 543 & 49.72 & 32.97 & 17.31 \\
        Time-Based Simulation & 510 & 53.53 & 33.14 & 13.33 \\
        Database Repository & 981 & 58.92 & 28.95 & 12.13 \\
        Database Query & 661 & 46.75 & 34.95 & 18.31 \\
        Exception Handling & 665 & 52.93 & 33.38 & 12.78 \\
        Version Dependency Issues & 1015 & 47.59 & \textbf{40.59} & 11.82 \\
        Errors in PHP Libraries & 691 & 51.37 & 34.01 & 14.62 \\
        Errors in Mocking Frameworks & 652 & 51.84 & 33.28 & 14.88 \\
        Errors in Android & 930 & 49.14 & 30.22 & \textbf{20.65} \\
        Use Case of Frameworks & 546 & 66.3 & 23.81 & 9.89 \\
        Frameworks Selection & 689 & \textbf{66.62} & 26.85 & 6.53 \\
        Tool Setup Errors & 489 & 45.81 & 35.38 & 18.81 \\
        Mocking Strategies & 1007 & 60.48 & 28.5 & 11.02 \\
        \bottomrule
    \end{tabular}
\end{table*}
\begin{table*}
    \centering
    \caption{Comparison of Popularity and Difficulty with other SE Fields}
    \label{tab:fields_stats}
    \small
    \begin{tabular}{l|l|l|l|l|l|l|l|l}
        \toprule
        & Metrics & Mocking & IoT & Big Data & Chatbot & Security & Mobile & Concurrency \\
        \midrule
        \midrule
        \multirow{3}{*}{Popularity} 
         & \# Posts & 25,302 & 53,173 & 125,671 & 3,890 & 94,541 & 1,604,483 & 245,541 \\
         & Avg View & 4,696.46 & 1,320.3 & 1,560.4 & 512.4 & 2,461.1 & 2,300.0 & 1,641 \\
         & Avg Score & 4.42 & 0.8 & 1.4 & 0.7 & 2.7 & 2.1 & 2.5 \\
        \midrule
        \midrule
        \multirow{2}{*}{Difficulty} 
         & \% W/O Acc Ans & 45.41\% & 64\% & 60.3\% & 67.7\% & 48.2\% & 52\% & 43.8\% \\
         & Med Hrs to Acc. & 1.28 & 2.9 & 3.3 & 14.8 & 0.9 & 0.7 & 0.7 \\
        \bottomrule
    \end{tabular}
\end{table*}
\begin{table}
\centering
\caption{Performance Metrics for \textit{GPT-4o-mini} model}
\begin{tabular}{lll}
\hline
\textbf{Metric} & \textbf{Macro (\%)} & \textbf{Micro (\%)} \\
\hline
Precision        & 79.01 & 88.78 \\
Recall           & 60.42 & 88.78 \\
F1 Score         & 62.84 & 88.78 \\
\hline
\end{tabular}
\label{tab:get4mini_metrics}
\end{table}
\begin{tcolorbox}
\textbf{RQ3:} \textit{Mocking}-related questions are notably more popular than the average SO questions, with a normalized popularity score of \textit{1.17}. Our analysis reveals an inverse relationship between the popularity and difficulty of mocking-related topics. Highly popular topics, such as \textit{Framework Selection}, exhibit high question scores and engagement while having lower difficulty, with only 33.38\% of questions lacking accepted answers. In contrast, challenging topics like \textit{HTTP Requests and Responses} and \textit{JS Modules and Components} have high unanswered rates (58.85\% and 58.7\%) and longer resolution times (6 and 7 hours, respectively).
\end{tcolorbox}
\textbf{RQ4: What types of questions do developers ask about mocking?}\\
\textbf{Motivation:} Understanding how developers use mocking techniques in testing is essential to identify challenges and best practices. Previous studies show that developers ask different types of questions (how, why, what) when encountering issues with a framework~\cite{treude2011programmers}. Categorizing these questions can help uncover the nature of difficulties faced during mock testing. 
\begin{table*}
    \centering
    \caption{Comparison of Question Types among Mocking Categories}
    \label{tab:compare_category_questiontypes}
    \small
    \begin{tabular}{l|l|l|l|l|l}
        \toprule
        Question Type & Mocking Techniques & Advanced Programming & External Services & Error Handling  & Theoretical \\
        \midrule
        How (\%) & 72.92 & 76.33 & 79.46 & 54.58 & 64.59 \\
        \midrule
        Why (\%) & 25.03 & 20.83 & 17.91 & 42.59 & 22.63 \\
        \midrule
        What (\%) & 1.86 & 2.51 & 2.34 & 2.50 & 10.80 \\
        \midrule
        Others (\%) & 0.19 & 0.33 & 0.28 & 0.28 & 1.98 \\
        \bottomrule
    \end{tabular}
\end{table*}
\begin{table*}
    \centering
    \caption{Comparison of Question Types among Mocking Topics}
    \label{tab:mocking_question_types}
    \small
    \begin{tabular}{lllll}
        \toprule
        \textbf{Topic} & \textbf{How \%} & \textbf{Why \%} & \textbf{What \%} & \textbf{Other \%} \\
        \midrule
        Configure Return Values & 57.02 & 41.81 & 0.97 & 0.19 \\
        Setup Method Parameter & 69.43 & 28.86 & 1.62 & 0.09 \\
        Argument Matching & 75.63 & 22.10 & 2.03 & 0.24 \\
        Ruby Methods Mocking & 69.19 & 20.93 & 9.69 & 0.19 \\
        Advanced Method Mocking & 83.18 & 14.95 & 1.62 & 0.24 \\
        Function Calls with Dependencies & 84.55 & 14.42 & 1.03 & 0.01 \\
        Mocking Python Components & 73.27 & 25.35 & 1.32 & 0.06 \\
        JS Modules and Components & 74.26 & 24.41 & 1.33 & 0.01 \\
        Objects and Properties & 81.45 & 16.91 & 1.34 & 0.30 \\
        Handling OOP Issues & 82.45 & 14.63 & 2.66 & 0.13 \\
        Interfaces and Generics & 74.85 & 19.19 & 5.67 & 0.29 \\
        Mocking Class Constructor & 83.41 & 14.02 & 2.32 & 0.12 \\
        Controller and HTTP Contexts & 75.29 & 24.14 & 0.57 & 0.01 \\
        Method Call Verification & 67.68 & 28.56 & 3.58 & 0.18 \\
        Web Services and APIs & 86.10 & 9.95 & 3.79 & 0.16 \\
        HTTP Requests and Responses & 75.79 & 21.82 & 2.39 & 0.01 \\
        File and Directory Access & 86.67 & 11.85 & 1.11 & 0.37 \\
        Device Dependent Events & 78.45 & 18.60 & 2.58 & 0.37 \\
        Time-Based Simulation & 78.04 & 20.20 & 1.57 & 0.20 \\
        Database Repository & 72.99 & 25.89 & 0.71 & 0.41 \\
        Database Query & 85.78 & 10.29 & 3.03 & 0.91 \\
        Exception Handling & 57.06 & 41.44 & 0.30 & 1.20 \\
        Version Dependency Issues & 57.14 & 37.73 & 5.02 & 0.10 \\
        Errors in PHP Libraries & 65.12 & 33.29 & 1.45 & 0.01 \\
        Errors in Mocking Frameworks & 56.29 & 39.57 & 4.14 & 0.01 \\
        Errors in Android & 41.08 & 57.96 & 0.97 & 0.01 \\
        Use Case of Frameworks & 63.55 & 28.21 & 7.69 & 0.37 \\
        Frameworks Selection & 58.93 & 14.37 & 24.24 & 2.47 \\
        Tool Setup Errors & 47.24 & 47.85 & 4.50 & 0.41 \\
        Mocking Strategies & 78.05 & 13.51 & 6.45 & 1.99 \\
        \bottomrule
    \end{tabular}
\end{table*}

\textbf{Approach:} Following prior research \cite{rosen2016mobile}, we classify \textsc{StackOverflow} posts related to \textit{mocking} into four distinct categories: \textit{How}, \textit{Why}, \textit{What}, and \textit{Other}. These categories capture different types of queries that developers make while working with mocking techniques. As a first step, we preprocess the text by removing unnecessary symbols, noise, and extra spaces. We then concatenate the title and body of each post and use the \textit{GPT-4o-mini} model for automated labeling. This classification was performed using an instruction-based prompt engineering approach, in which the model was provided with categorical definitions to perform a zero-shot classification. The prompt was carefully designed to specify the model’s role, and constrain its output to predefined labels (\textit{How}, \textit{Why}, \textit{What}, and \textit{Other}), thereby facilitating reliable and scalable annotation of the dataset without the need for fine-tuning.

To assess the performance of the model, the first and second authors manually labeled a subset of 500 randomly selected posts. We then compared the model's predictions against these labels, evaluating its performance using precision, recall, and F1 scores as shown in Table~\ref{tab:get4mini_metrics}. The \textit{GPT-4o-mini} model achieved strong micro-level performance across all metrics \textit{88.78\%}, indicating consistent effectiveness across individual predictions. At the macro level, which reflects performance across different classes, the model obtained \textit{79.01\%} precision, \textit{60.42\%} recall, and \textit{62.84\%} F1 score. Based on these results, we used the \textit{GPT-4o-mini} model to automatically classify the entire dataset of 25,302 posts. This approach allows for efficient large-scale classification while maintaining reliability through human validation. 

After obtaining the labeled dataset as output, we calculate the percentage distribution of each category to identify which types of questions are most common among users. The four question types are outlined as below:

\textbf{How:} Posts in this category seek guidance on implementing a specific method, technique, or approach related to Mocking. These questions focus on achieving a particular goal and request step-by-step instructions or best practices. For example, \href{https://stackoverflow.com/q/60464234}{Q60464234} which asks “How to mock a file with no read permission in Python3?”. 

\textbf{Why:} This category includes posts where developers seek explanations for unexpected behavior, errors, or underlying causes of issues. These questions are often related to troubleshooting and debugging. For example, \href{https://stackoverflow.com/q/14858286}{Q37457756} asks “Why phpunit doesn't run \texttt{\_\_destruct()} in mocked class and how to force it?”.

\textbf{What:} Posts that fall under this category request specific information, definitions, or clarifications about mocking concepts. Developers typically ask these questions to gain a better understanding before making informed decisions. For example, \href{https://stackoverflow.com/q/37457756}{Q37457756}, which asks “What is the equivalent of mockito's spy in jmock 2?”.

\textbf{Other:} This category includes posts that do not fit into the above classifications. These may involve open-ended discussions, opinions, or multiple unrelated questions. For example, \href{https://stackoverflow.com/q/49325188}{Q49325188} asks discusses about “Mocking Spring AOP Aspects and Handling NullPointerException”.

\textbf{Results:} Table \ref{tab:compare_category_questiontypes} presents the comparison of question types across different mocking categories. Most of the questions in all categories focus on \textit{How} (ranging from 54.58\% in Error Handling to 79.46\% in External Services), indicating a strong emphasis on practical implementation. \textit{Why} questions are the second most common, particularly in Error Handling (42.59\%), suggesting a need for a deeper understanding of underlying principles in this category. \textit{What} questions remain relatively low, with the highest occurrence in Theoretical (10.80\%), likely reflecting a conceptual focus. \textit{Others} account for a negligible proportion in all categories, confirming that most inquiries fall within \textit{How} and \textit{Why} classifications.

Next, we demonstrate a comparison on the distribution of question types across various mocking topics in Table \ref{tab:mocking_analysis}. Across most topics, \textit{How} questions dominate, particularly in areas like \textit{File and Directory Access} (86.67\%), \textit{Web Services and APIs} (86.10\%), and \textit{Database Query} (85.78\%), reflecting a strong emphasis on practical implementation of External Services. \textit{Why} questions are the second most common overall and are especially prevalent in topics like \textit{Errors in Android} (57.96\%), \textit{Tool Setup Errors} (47.85\%), and \textit{Exception Handling} (41.44\%), indicating a need for deeper conceptual understanding in error-related discussions. \textit{What} questions occur infrequently, but are relatively higher in \textit{Frameworks Selection} (24.24\%) and \textit{Ruby Methods Mocking} (9.69\%), suggesting more conceptual inquiries in these areas. 

\textbf{Comparison with other SE Fields:} Next, we examine how user concerns in the mocking domain differ from those in the IoT, Mobile, and Chatbot domains. To perform this analysis, we compare the distribution of question types\textemdash{}classified as \textit{How}, \textit{Why}, \textit{What}, and \textit{Others}\textemdash{}across domains. However, such detailed classification results have not been consistently reported in all previous studies of \textsc{StackOverflow} data. Specifically, for many SE domains (e.g., Big Data, Security, Concurrency) existing literature does not include a breakdown of question types using this classification scheme. As a result, we limit our comparison to the three domains (IoT, Mobile, and Chatbot) for which this information is available.

Table~\ref{tab:compare_questiontypes} compares the types of questions across different SE fields, showing that mocking has the highest percentage of \textit{How} questions (69.58\%), indicating a practical, implementation-focused nature compared to IoT (47.3\%), Chatbot (61.8\%), and Mobile (59\%). Mocking also has a relatively high proportion of \textit{What} questions (25.80\%), higher than Chatbot (11.7\%) and Mobile (8\%), but lower than IoT (37.9\%). This suggests that, developers frequently seek to understand the underlying principles or terminology associated with mocking\textemdash{}such as tools, behaviors, or framework-specific features\textemdash{}before applying them in practice. \textit{Why} questions are less common in Mocking (4\%), suggesting fewer diagnostic or error-focused inquiries compared to Mobile (29\%) and Chatbot (25.4\%).
\begin{table}[h]
    \centering
    \caption{Comparison of Question Types with other SE Fields}
    \label{tab:compare_questiontypes}
    \small
    \begin{tabular}{l|c|c|c|c}
        \toprule
        Question Type & Mocking & IoT & Chatbot & Mobile \\
        \midrule
        How (\%) & \textbf{69.58\%} & 47.3\% & 61.8\% & 59\% \\
        \midrule
        What (\%) & 25.80\% & \textbf{37.9\%} & 11.7\% & 8\% \\
        \midrule
        Why (\%) & 4.00\% & 20\% & 25.4\% & \textbf{29\%} \\
        \midrule
        Others (\%) & 0.61 \% & \textbf{8.3\%} & 1.2\% & 7\% \\
        \bottomrule
    \end{tabular}
\end{table}
\begin{tcolorbox}
\textbf{RQ4:} Most developer questions about \textit{Mocking} focus on \textit{How} to implement techniques (over 70\%), followed by \textit{Why} questions in error-related contexts, highlighting debugging concerns. \textit{What} questions are rare, peaking in theoretical discussions. \textit{How} questions dominate, particularly in areas like \textit{External Services} that include topics such as \textit{File and Directory Access}, \textit{Web Services and APIs}, and \textit{Database Query}. Compared to other SE domains such as IoT, Mobile, and Chatbot, \textit{Mocking} exhibits the highest proportion of \textit{How} questions, reinforcing its implementation-centric nature.
\end{tcolorbox}

\section{Implications}
\label{sec:discussion}
The findings from our study provide important implications for software developers, educators, and researchers in the field of Test Mocking.

\textbf{Implications for Software Developers.}
Developers need to pay careful attention to documentation and community discussions when troubleshooting issues with Mocking Techniques~\cite{ibarra2023mock, kochhar2019practitioners}. The findings from our study provide valuable insights into the challenges and areas of improvement in mocking frameworks for software developers . One key implication is the need to be proactive in selecting the appropriate mocking tools for specific scenarios. 
\begin{figure*}[t] 
    \centering
    \includegraphics[width=\textwidth,keepaspectratio]{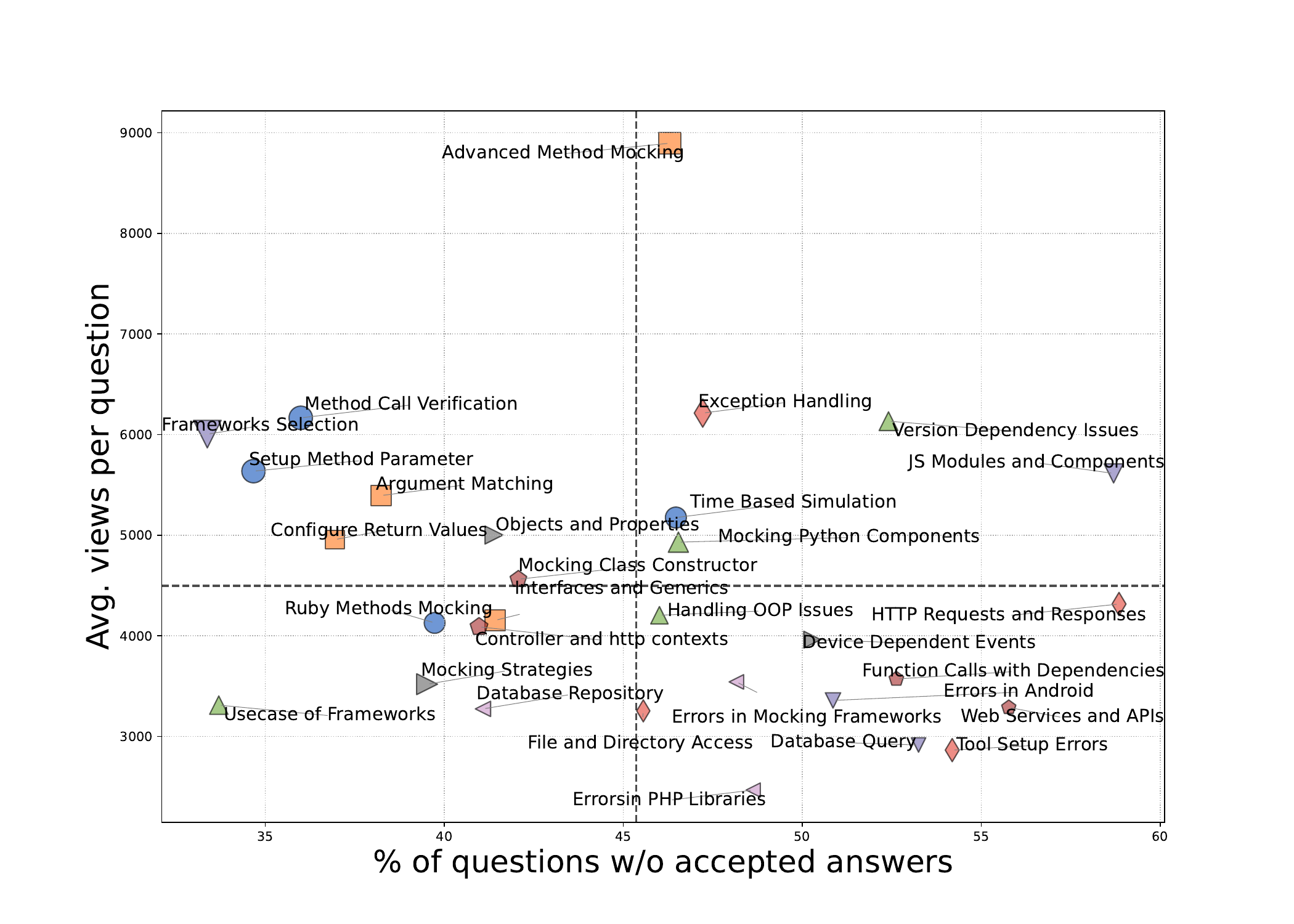} 
    \caption{Tradeoff between Question Popularity (Avg. views) and Difficulty (Answer Acceptance)}
    \label{fig:bubble_chart}
\end{figure*}

The bubble chart in Figure~\ref{fig:bubble_chart} illustrates the tradeoff between question popularity (views) and difficulty (answer acceptance), providing valuable insights for software developers. Highly viewed but poorly answered topics, such as \textit{Advanced Method Mocking} and \textit{JS Modules and Components} suggest areas of complexity where developers often struggle to find clear solutions, making them crucial for deeper learning and knowledge sharing. In contrast, well-answered yet popular topics, like \textit{Method Call Verification} indicate common challenges with widely available solutions, making them more approachable. Topics with low views and low answer acceptance, such as \textit{Errors in Android} and \textit{Tool Setup Errors} may indicate niche or unresolved problems, highlighting potential research or documentation gaps. Understanding these trends can help developers prioritize learning, improve problem-solving efficiency, and contribute meaningfully to the developer community by addressing gaps in knowledge.

\textbf{Implications for Educators.} For educators, the results of this study highlight the importance of providing clear and accessible resources to help students and junior developers navigate the complexities of mocking frameworks. The prominence of \textit{Framework Selection} in the discussions on \textsc{StackOverflow} suggests that this is crucial topic that should be emphasized in curricula. Educators should aim to cover the practical aspects of mocking, including how to choose the right tool for different testing scenarios, as well as how to properly implement mocking in unit tests~\cite{schroeder2005teaching}.

Additionally, as the study shows a significant focus on How and What questions, educators should prioritize developing educational materials and tutorials that address these types of questions. By providing step-by-step guides, FAQs, and other practical resources, educators can empower students to become more self-sufficient in solving real-world testing challenges~\cite{ihantola2010review}. There is also an opportunity to develop video tutorials or interactive learning platforms that allow students to experiment with mocking frameworks in a controlled environment, encouraging active learning and deeper understanding.

\textbf{Implications for Researchers.} This study points to the importance of understanding the evolving needs of the developer community. Researchers could investigate the effectiveness of different mocking strategies for specific testing scenarios and propose improvements to existing frameworks, such as incorporating automated recommendations for best practices during testing~\cite{wang2021automatic}.

Researchers could develop automated tools to support developers during mocking. Features like auto-completion for mock objects, intelligent suggestions for best practices, and detection of misconfigurations can reduce errors and improve test reliability. Tools that generate mocks based on real dependencies and ensure consistency could streamline workflows and encourage proper mocking practices, ultimately leading to more robust and maintainable codebases

Additionally, researchers could explore the gaps between theoretical and practical knowledge in mocking frameworks, particularly in terms of how developers transition from understanding the concepts of mocking to applying them effectively in real-world projects. This could involve empirical studies or case studies to identify common stumbling blocks for developers and propose solutions that improve the adoption of mocking techniques in production environments.
\section{Threats To Validity} 
\label{sec:threats}
In this section, we discuss the potential threats to the validity of our study. These threats are categorized into internal, external, and construct validity. We outline how we mitigated each threat and how it could still impact the conclusions of the study.

\textbf{Internal Validity:} Our study relied on the application of Latent Dirichlet Allocation (LDA) for topic modeling that introduces internal threats. The choice of the optimal number of topics (K) is crucial for the accuracy of the results. In our study, we experimented with K values from 1 to 100 and selected the value that maximized topic coherence~\cite{stevens2012exploring}. However, it is possible that the true optimal number of topics could be different, as topic modeling inherently involves some degree of subjectivity~\cite{chang2009reading}. We minimized this risk by using a well-established approach, which included visualizing the topics with LDAvis~\cite{sievert2014ldavis} and selecting K values that produced distinct and interpretable topics.

The process of labeling topics based on the top 10 words generated by LDA and top 30 relevant posts is subject to human interpretation and bias. While we used the open card sorting method with two researchers to label and refine the topics, the subjective nature of this process could introduce inconsistency. To reduce this threat, we conducted several iterations of the card-sorting process, ensuring mutual agreement and refining the labels until they were robust and meaningful.

Another internal threat in our approach is the potential misclassification of mocking-related \textsc{StackOverflow} posts due to the reliance on GPT for automated labeling. GPT may introduce biases or errors, especially for nuanced or ambiguous posts that do not fit clearly into predefined categories. To mitigate this, we performed manual validation on a random subset of 500 posts and measured the accuracy of the GPT model on the random samples. Although this helps to assess the reliability of the automated classification, we acknowledge that it may not completely eliminate errors.

\textbf{External Validity:} Our study relied solely on \textsc{StackOverflow} as the source of data for mocking-related discussions. While \textsc{StackOverflow} is a popular and widely used platform among developers, it may not fully represent the broader developer community. Developers who do not use \textsc{StackOverflow} or prefer other forums may face different challenges and discuss other aspects of mocking. Additionally, the topics discussed on \textsc{StackOverflow} may not be entirely representative of the industry as a whole. However, we recognize that incorporating data from additional forums or surveys could enhance the external validity of the study.

\textbf{Construct Validity:} Construct validity refers to the degree to which our study measures what it intends to measure. We assumed that LDA’s probabilistic topic modeling approach adequately captures the underlying topics in the dataset. Although LDA is a well-established method in software engineering research, it has its limitations, including the inherent randomness of its output. To address this, we performed multiple runs of the LDA model for different iterations and evaluated the coherence scores to ensure that our results were consistent.

Recently, large language model (LLM)-based approaches such as BERTopic have gained popularity for topic modeling tasks across various domains, including software engineering~\cite{azher2024limtopic, bu2023software, naghshzan2023enhancing}. These models leverage contextual embeddings from transformer-based architectures to capture semantic meaning more effectively than traditional methods. However, despite their promise, LLM-based models are not without drawbacks. Studies have shown that BERTopic's reliance on pre-trained language models may not generalize well to technical domains like software engineering due to the presence of code, domain-specific jargon, and mixed-format text~\cite{azher2024limtopic, zhang2023unifying}.

Additionally, BERTopic's use of clustering and dimensionality reduction introduces variability in results, making reproducibility a concern~\cite{grootendorst2022bertopic}. Furthermore, there is still no established standard for applying BERTopic in software engineering research, whereas Latent Dirichlet Allocation (LDA) has a long-standing history of successful applications in the field~\cite{bajaj2014mining, barua2014developers, yang2016security, rosen2016mobile}. Given these considerations, we opted to use LDA for topic modeling in our study to ensure interpretability, consistency, and alignment with prior work in the domain. However, future work can explore BERTopic-based topic modeling to analyze Stack Overflow data.

Furthermore, the analysis of question popularity and difficulty is subject to measurement limitations. For instance, we used upvotes and answer counts to determine popularity and time to get accepted answers to measure difficulty. However, these metrics may not fully capture the nuances of a post’s popularity or the complexity of the issue being discussed. Not all developers mark answers as accepted, and upvotes alone may not always reflect the actual helpfulness of the answer. These metrics can be insufficient for accurately assessing question quality and engagement in platforms like \textsc{StackOverflow}~\cite{treude2011programmers, shah2010evaluating}.

\section{Conclusion}
\label{sec:conclusion}
Our study analyzed \textsc{StackOverflow} discussions on \textit{Mocking}, revealing key trends in developer concerns and challenges. We identified 5 major categories, 13 sub-categories, and 30 specific topics. \textit{Mocking Techniques} is the most discussed category, particularly topics such as \textit{Mocking Python Components} and \textit{JS Modules and Components}. Over time, \textit{Mocking Techniques} consistently remained dominant, while \textit{Theoretical} discussions about mocking diminished, signaling a shift toward practical concerns. We observed that \textit{Mocking}-related topics vary in popularity and difficulty, with \textit{Framework Selection} being a widely discussed topic, while more complex areas like debugging \textit{HTTP Requests and Responses} and \textit{Web Services and APIs} presented significant challenges. Popular topics often received quicker resolutions, while more complex issues had longer response times. Developers primarily ask \textit{How}-type questions about \textit{External Services}, reflecting the challenges related to integration and practical implementation. \textit{Why}-type questions were more common in \textit{Error Handling}, highlighting struggles with debugging and understanding underlying principles. Our findings emphasize the practical nature of mocking-related discussions, with a focus on solving integration issues and debugging complex scenarios. These insights highlight the evolving nature of mock testing discussions, suggesting a growing need for better tooling, resources, and documentation to support developers. Future research could explore how emerging frameworks and practices impact mocking in the industry and how they shape developers' approach of \textit{Mocking} in real-world applications.
\section*{Acknowledgment}
This research is supported in part by the Natural Sciences and Engineering Research Council of Canada (NSERC) Discovery Grants program, the Canada Foundation for Innovation's John R. Evans Leaders Fund (CFI-JELF), and by the industry-stream NSERC CREATE in Software Analytics Research (SOAR). We also gratefully acknowledge the contributions of students and faculty members from the Department of Computer Science at the University of Saskatchewan and the University of Manitoba.




\bibliographystyle{elsarticle-num}

\bibliography{paper}





\end{document}